\documentclass[twocolumn,prb,aps,floatfix,superscriptaddress]{revtex4-2}
\usepackage[latin9]{inputenc}
\usepackage{amsmath}
\usepackage{amssymb}
\usepackage{graphicx}

\makeatletter
\usepackage{bm}\usepackage{epsfig}\usepackage{color}\usepackage{siunitx}
\usepackage{subfigure}

\makeatother

\begin{document}
\title{Coulomb interactions and renormalization of semi-Dirac fermions near
a topological Lifshitz transition}
\author{Valeri N. Kotov}
\affiliation{Department of Physics, University of Vermont, Burlington, VT 05405}
\author{Bruno Uchoa}
\affiliation{Center for Quantum Research and Technology, Department of Physics
and Astronomy, University of Oklahoma, Norman, OK 73069}
\author{Oleg P. Sushkov}
\affiliation{School of Physics, University of New South Wales, Sydney 2052, Australia}
\date{\today}
\begin{abstract}
 We aim to understand how the spectrum of semi-Dirac fermions
is renormalized due to long-range Coulomb electron-electron interactions
at a topological Lifshitz transition, where two Dirac cones merge.
At the transition, the electronic spectrum is characterized by massive
quadratic dispersion in one direction, while it remains linear in
the other. We have found that, to lowest order, the unconventional
log squared (double logarithmic) correction  to the quasiparticle mass 
in bare perturbation theory leads to resummation into  strong mass renormalization 
in the exact full solution of the perturbative renormalization group
equations. This behavior effectively wipes out the curvature of the
dispersion and leads to  Dirac cone restoration at low energy: the
system flows towards  Dirac dispersion which  is anisotropic but linear
in momentum, with interaction-depended logarithmic modulation. The Berry phase associated 
with the restored critical  Dirac spectrum  is zero - a property guaranteed  by time-reversal symmetry and 
unchanged by renormalization. Our results are in contrast with the behavior that has been found within the
large-$N$ approach.
\end{abstract}
\maketitle

\section{Introduction}

Semi-Dirac fermions are chiral quasiparticles in two dimensions (2D) that propagate
as Galilean invariant particles as they move in one direction and
as relativistic ones in the other direction. Such quasiparticles emerge
at a topological Lifshitz transition, where two Dirac cones merge
\cite{Montambaux2009,Montambaux2009-2,Montambaux2016,Bellec2013,Lim2012,Banarjee2009,Banarjee2012}.
Strongly anisotropic Dirac fermions, eventually transforming into
semi-Dirac particles at a topological quantum critical point, appear
in a variety of physical situations, from strained graphene-based
structures \cite{Maria2016}, black phosphorus under pressure \cite{PhysRevLett.112.176801}
and doping \cite{Kim723}, BEDT-TTF$_{2}$I$_{3}$ salt under pressure \cite{doi:10.1143/JPSJ.75.054705},
VO$_{2}$/TO$_{2}$ heterostructures \cite{PhysRevLett.102.166803, PhysRevB.92.161115},
photonic crystals and atomic (cold atom) physics \cite{Photonic2013,Polini2013}.
In solid state context the prototypical example is strained graphene.
It is known that by applying uniaxial strain in the the zig-zag direction
in graphene one can induce a transition into a gapped state. In the
gapless regime (before the transition), the electronic spectrum consists
of separated anisotropic (elliptic) Dirac cones, while at the transition
the spectrum becomes quadratic in one direction, remaining linear
in the other \cite{Pereira2009,Montambaux2009,Montambaux2009-2,Choi2010}.


\begin{figure}[b]
\begin{centering}
\includegraphics[scale=0.28]{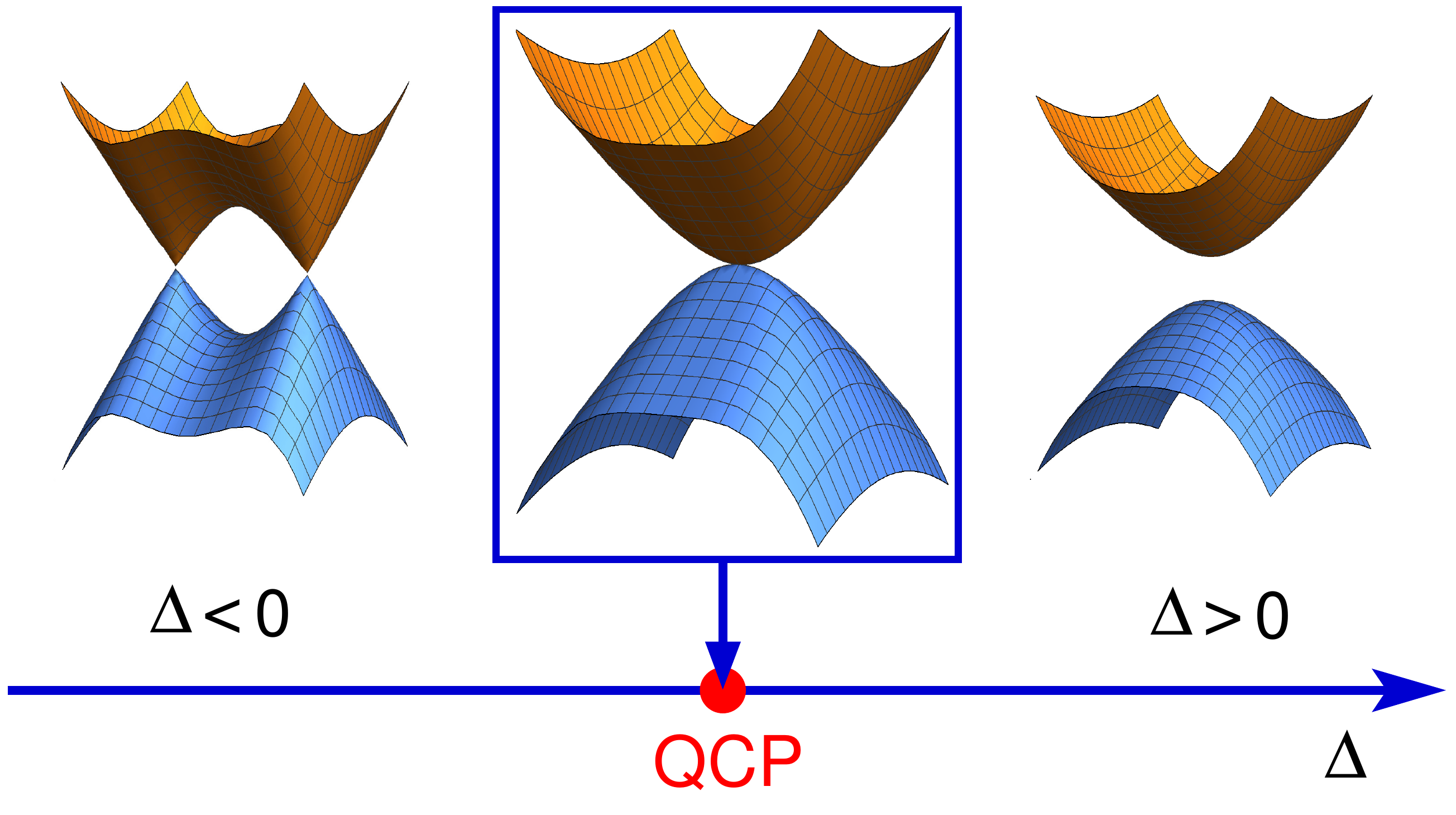}
\par\end{centering}
\caption{{\small{}\label{fig:QCP}Topological Lifshitz phase transition across a quantum critical
point (QCP) at $\Delta=0$. For $\Delta<0$, approaching the QCP from
the left, two Dirac cones merge, producing a single touching point
with semi-Dirac fermion excitations. For $\Delta>0$, a trivial insulating
phase forms.}}
\end{figure}


The universal effective Hamiltonian describing the physics outlined
above is 
\begin{equation}
\mathcal{H}(\mathbf{p})=\left(\frac{p_{x}^{2}}{2m}+\Delta\right)\hat{\sigma}_{x}+vp_{y}\hat{\sigma}_{y},\label{ham}
\end{equation}
where $\Delta$ depends on the (anisotropic) hopping parameters, in
the case of strained graphene.  We will keep in mind this example,
while the results will be of course applicable to all systems falling
within the same universality class. The case $\Delta<0$ corresponds
to separated anisotropic (elliptic) Dirac cones (gapless phase, weak
strain), the value $\Delta=0$ is the critical point, and $\Delta>0$
corresponds to the gapped phase (strong strain), as shown in Fig.~\ref{fig:QCP}.
The chemical potential is set to zero. 

At the critical point the spectrum is
\begin{equation}
\varepsilon({\bf p})=\pm\sqrt{\left(\frac{p_{x}^{2}}{2m}\right)^{2}+v^{2}p_{y}^{2}},\ \ \ \ \Delta=0.\label{semi-dirac-dispersion}
\end{equation}
From now on we set $\hbar=1$ and all lengths will be measured in
units of the lattice spacing (which we set to one), with $\pm$ indexing
the two particle-hole branches. In particular, at the critical point
induced by zig-zag strain, by taking into account the strain dependence of the tight-binding Hamiltonian parameters,
 one can deduce the following relationship, $mv=2$,
in units of the inverse lattice spacing \cite{Pereira2009,Montambaux2009}. This is the only remnant
of non-universal (system specific) physics at the critical point and
we use it for illustration purposes in our plots describing interaction
effects (whose structure itself is universal.)

An important issue is how interactions (both short and long-range)
affect the fermion spectrum at and around the critical point, and
the various phenomena associated with it. For example short range
interactions can influence the Dirac cone merger and shift the critical
point itself (i.e. affect the gap) \cite{Dora2013}. Such interactions
can also affect the appearance of various instabilities (such as charge
and spin density waves, etc) at criticality \cite{Uryszek2019,PhysRevB.96.220503,Foster2018}.

The role of long-range Coulomb interactions is expected to be even
more dramatic. It has been argued \cite{Isobe2016,Cho2016,Link2018}
that a non-Fermi liquid (NFL) state emerges in the large $N$ limit, where
the quasiparticle residue ($Z$) approaches zero as a power law at low energy. This behavior 
is governed by the $N\alpha \gg 1$ limit. 
At the lowest energies, this state crosses over to a marginal Fermi liquid (MFL)
where $Z$ exhibits a weaker logarithmic renormalization, governed by 
a weak coupling (in a sense that $N\alpha \ll 1$) fixed point.
Here
$N$ is the number of fermion flavors (equal to four) and $\alpha$ is the effective Coulomb 
coupling constant. 
This overall behavior  can be
compared with previous results for simple, isotropic Dirac cones in
graphene within the same approximation \cite{Son2007,Kotov2009,Kotov2012}
where $Z$ does not vanish and the interacting isotropic Dirac liquid
remains coherent. The peculiar incoherent behavior of the semi-Dirac
fermions can be traced back to the appearance of higher powers of
logarithms in perturbation theory (log squared contributions even
at first order of perturbation theory, compared to simple logs for
isotropic graphene). It should be emphasized that this result is based
on the large $N$ scheme, i.e. assuming the dominance of polarization
bubbles. The alternative to large $N$ is the ``conventional" perturbative
renormalization group (RG) in powers of the Coulomb coupling $\alpha$.
While in isotropic graphene the two approaches connect smoothly and
describe the same state (interacting Dirac liquid) \cite{Kotov2012},
for semi-Dirac fermions the results are drastically different, as
we will show below.

The purpose of the present paper is to point out that for semi-Dirac fermions the 
``NFL--MFL" fixed point obtained in the large $N$ limit is not the only
possible scenario. The presence of log squared terms in first order
of perturbation theory does not by itself justify non-perturbative
RG when $\alpha$ is small and $N\sim1$. We show
that after taking into account the unconventional log squared contributions
that appear in the self-energy for semi-Dirac fermions, and performing
perturbative RG to lowest order in $\alpha$, the resulting fixed point is characterized by restoration
of linear quasiparticle dispersion in the direction where it was originally
quadratic. The resulting Dirac cone is not necessarily isotropic but
the ``semi-Diracness" has disappeared.  We also emphasize that even though the interaction effects tend to restore   the linear Dirac dispersion,
 the Berry phase, which is zero for the bare semi-Dirac Hamiltonian \cite{Montambaux2009}, remains zero upon renormalization.  
 The zero value of the Berry phase is a topological
 property which is guaranteed  by the fact that the semi-Dirac spectrum is a result of a merger of two Dirac
 cones (related by time-reversal symmetry) with Berry phases $\pm \pi$.
The behavior we find is  in contrast to the MFL state where
the dispersion retains its semi-Dirac features \cite{Isobe2016}. While we have
not addressed the issue how the quasiparticle residue behaves, since it appears at the next order
in $\alpha$, we do not expect our main conclusion about Dirac cone restoration to be altered 
due to the fact that the residue affects the terms in the different momentum directions in the same manner.
Thus our  results indicate that
the perturbative RG and the large $N$ version lead to different fixed
points, and this can have far-reaching consequences for properties
of interacting semi-Dirac fermions.





For instance it has been claimed \cite{Link2018} that, at large $N$,
the ratio between the shear viscosity and the entropy of semi-Dirac
fermions violates the conjectured lower bound $\eta/s\geq\hbar/(4\pi k_{B})$
derived in an infinitely strongly coupled conformal field theory \cite{PhysRevLett.94.111601}.
This ratio is usually taken as a universal measure of the strength
of interactions in the hydrodynamic regime of quantum fluids. The
violation was attributed to the strongly anisotropic nature
of semi-Dirac fermions \cite{Link2018}. In contrast, conventional Dirac fermions are
known to satisfy the lower bound \cite{PhysRevLett.103.025301}. In
the present work we find that, at least in the perturbative regime,
Coulomb interactions lead to  restoration
of the linearity of the spectrum. This effect may have relevant implications
for  the solution of the quantum kinetic equation in the collision dominated
regime.

The rest of the paper is organized as follows. In Section II we present
a detailed formulation and results of the perturbative RG for semi-Dirac
fermions at criticality. In Section III we discuss  issues related to the self-consistency
of our approach  which include examination of screening at weak coupling.
Section IV contains implications of our results for physical observables.
In Section V we also extend our treatment
away from the critical point. Section VI contains our conclusions.

\section{Renormalization Group at Criticality: Restoration of Dirac Spectrum
at Low Energy}

\begin{figure}
\includegraphics[width=0.20\textwidth]{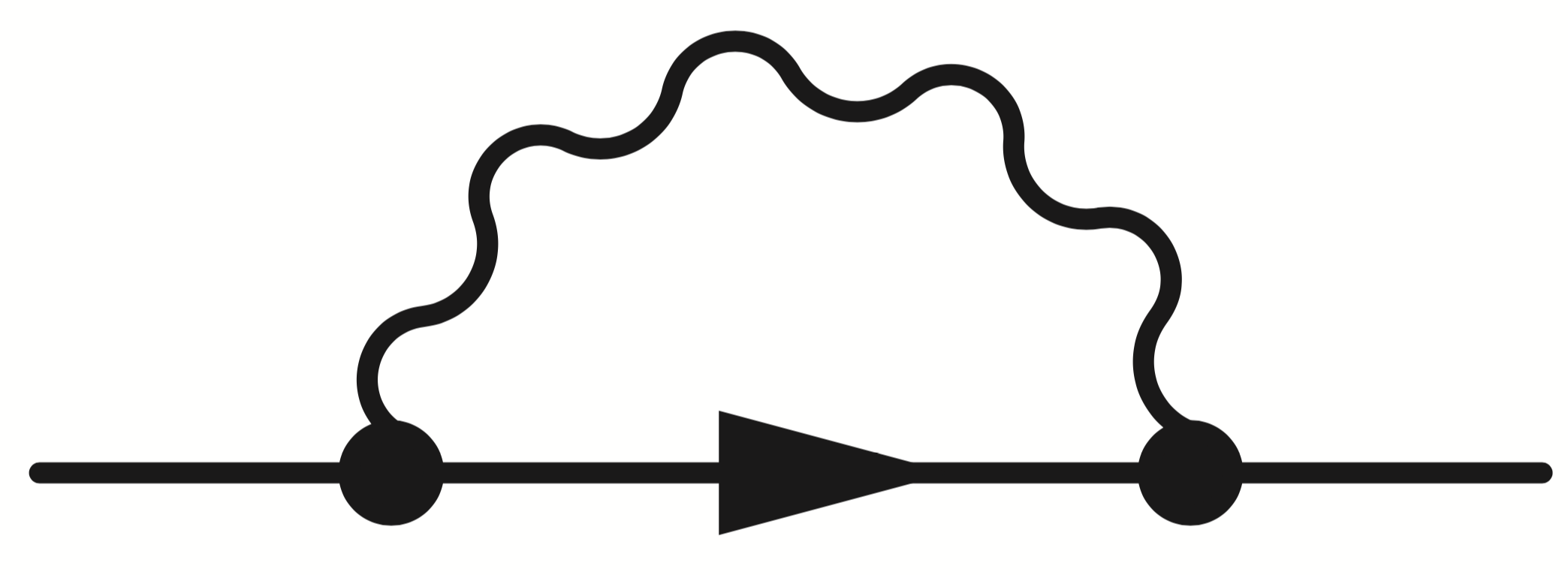} 
\caption{\label{fig:SE} Self-energy to first order in the Coulomb interaction
(wavy line).}
\end{figure}

In this section we consider the critical point $\Delta=0$. Let us
introduce interactions via the non-retarded Coulomb potential
\begin{equation}
V({\bf p})=\frac{2\pi e^{2}}{|{\bf p}|}.
\end{equation}
We will take into account the interaction at first order in perturbation
theory. The self-energy shown in Fig.~\ref{fig:SE} is
\begin{equation}
\hat{\Sigma}(\mathbf{p})=i\int_{-\infty}^{\infty}\frac{d\nu}{2\pi}\int\frac{d^{2}k}{(2\pi)^{2}}\hat{G}(\mathbf{k},\nu)V(\mathbf{k}-\mathbf{p}),\label{eq:Sigma0}
\end{equation}
where 
\begin{equation}
\hat{G}^{-1}(\mathbf{p},\nu)=\nu-\mathcal{H}(\mathbf{p}) + i0^{+}{\mbox{sign}}(\nu)\label{G}
\end{equation}
is the fermionic Green's function. The frequency integral can be easily
evaluated, 
\begin{equation}
\hat{\Sigma}({\bf p})=\frac{1}{2}\int\frac{d^{2}k}{(2\pi)^{2}}\frac{2\pi e^{2}}{|{\bf k}-{\bf p}|}\frac{1}{|\varepsilon({\bf k})|}\left(\frac{k_{x}^{2}}{2m}\hat{\sigma}_{x}+vk_{y}\hat{\sigma}_{y}\right).\label{se}
\end{equation}
In this order, the self-energy is frequency independent. When evaluating
logarithmic corrections it is useful to look at the behavior at small
external momenta $p\rightarrow0$ and expand
\begin{equation}
\frac{1}{|{\bf k}-{\bf p}|}=\frac{1}{k}\left\{ 1+\frac{{\bf k}.{\bf p}}{k^{2}}-\frac{{p}^{2}}{2k^{2}}+\frac{3({\bf k}.{\bf p})^{2}}{2k^{4}}\right\} +O(p^{3}).\label{p-expansion}
\end{equation}
Here $k=|{\bf k}|$. As usual, we introduce the dimensionless coupling
\begin{equation}
\label{coupling}
\alpha=e^{2}/v.
\end{equation}


\subsection{Gap Generation}

First we observe that, unlike the case of isotropic graphene, the
self-energy at zero momentum is finite,
\begin{equation}
\hat{\Sigma}({\bf p}=0)=\Delta_{0}\hat{\sigma}_{x},\label{Sigma2}
\end{equation}
implying that a gap is generated by the interactions. The mass gap
evaluated from Eq. (\ref{se}) is
\begin{equation}
\Delta_{0}=\alpha mv^{2}\int_{0}^{\tilde{\lambda}}kdk\int_{0}^{2\pi}\frac{d\varphi}{2\pi}\frac{\cos^{2}\varphi}{\sqrt{k^{2}\cos^{4}\varphi+\sin^{2}\varphi}},
\end{equation}
where $\tilde{\lambda}\equiv\lambda/2mv$ is the rescaled ultraviolet
cutoff, and $\lambda\sim1$ is the ultraviolet momentum cutoff in
units of the inverse lattice spacing (set to one in our convention).
To be specific, we evaluate this expression at the critical point
relevant to strained graphene, i.e. for $mv=2$ leading to $\tilde{\lambda}\approx1/4$.
At this point the integral that appears in the above equation is $0.046$.
The result is then $\Delta_{0}\approx0.1\alpha v$ (restoring the
units: $\Delta_{0}\approx0.1\alpha\hbar v/a$, where $a$ is the lattice
spacing).

Thus we can conclude that the interaction effects drive the system
away from criticality, towards the gapped phase ($\Delta_{0}>0$).
In the rest of this section we will assume that the system parameters
(for example anisotropic hopping parameters, strain, pressure, etc)
are externally fine tuned in such a way that the effective gap is
zero. This way we can study the spectrum renormalization at criticality.
We will return to the issue of gap renormalization in Section III.


\subsection{Mass and Velocity Renormalization}

We now proceed to calculate the first order corrections to the velocity
and mass parameters. These will exhibit logarithmic divergencies and
we will adopt an ``on-shell" renormalization procedure with an ultraviolet
energy cutoff $\Lambda$ which follows the structure of the dispersion
$\varepsilon({\bf p})$, and therefore depends on direction in momentum space.
 To extract the log divergence with an energy
cutoff we introduce a change of variables, 
\begin{equation}
\frac{k_{x}^{2}}{2m}=\varepsilon\sin\varphi,\ \ vk_{y}=\varepsilon\cos\varphi,\label{var-change}
\end{equation}
where $\varepsilon\in[0,\Lambda]$ and $\varphi\in[0,\pi]$. Then
we have 
\begin{equation}
\int d^{2}k=(1/v)\int_{0}^{\Lambda}\sqrt{2m\varepsilon}d\varepsilon\int_{0}^{\pi}d\varphi\frac{1}{\sqrt{\sin\varphi}}.\label{var-change-int}
\end{equation}
Integration over the variables $\varepsilon$ and $\varphi$ in the
self-energy (\ref{se}) gives
\begin{equation}
\hat{\Sigma}(\mathbf{p})=\left(\frac{p_{x}^{2}}{2m}\Sigma_{x}+\frac{p_{y}^{2}}{2m}\Sigma_{y,m}\right)\hat{\sigma}_{x}+vp_{y}\Sigma_{y,v}\hat{\sigma}_{y}.\label{Sigma3}
\end{equation}
The term 
\begin{equation}
\Sigma_{y,v}=\frac{\alpha}{4}\int_{E_{\omega}}^{E_{\Lambda}}\frac{dE}{E}L_{1}(E)=\frac{\alpha}{\pi}\ln{(\Lambda/\omega)},\label{Sigmayv}
\end{equation}
gives the self-energy correction to the velocity $v$, where
\[
L_{1}(E)=\int_{0}^{\pi}\frac{d\varphi}{\pi}\frac{E\cos^{2}\varphi}{\sqrt{\sin\varphi}\ (E\cos^{2}\varphi+\sin\varphi)^{3/2}}\stackrel{E\ll1}{\longrightarrow}\frac{4}{\pi}
\]
is an angular integral, and 
\begin{equation}
E=\varepsilon/\varepsilon_{0},\ \ \ \varepsilon_{0}\equiv2mv^{2},
\end{equation}
is the dimensionless energy integrated in the interval $E\in[E_{\omega},E_{\Lambda}]$,
with $E_{\omega}=\omega/\varepsilon_{0}$ and $E_{\Lambda}=\Lambda/\varepsilon_{0}$.
The renormalization is done ``on-shell" in the low-energy limit,
\begin{equation}
\omega\equiv|\varepsilon({\bf p})|=\sqrt{\left(\frac{p_{x}^{2}}{2m}\right)^{2}+v^{2}p_{y}^{2}}\ll\Lambda.\label{eq:omega}
\end{equation}

The first term in (\ref{Sigma3}) gives correction to the mass $m$
for quasiparticles moving along $p_{x}$, 
\begin{align}
\Sigma_{x} & =-\frac{\alpha}{8}\int_{E_{\omega}}^{E_{\Lambda}}\frac{dE}{E}L_{2}(E)+\frac{3\alpha}{8}\int_{E_{\omega}}^{E_{\Lambda}}\frac{dE}{R}L_{3}(E)\nonumber \\
 & =\frac{\alpha}{4\pi}\ln^{2}(\Lambda/\omega)+\frac{\alpha}{4\pi}F\ln(\Lambda/\omega),\label{Sigmax}
\end{align}
where 
\begin{align}
L_{2}(E) & =\int_{0}^{\pi}\frac{d\varphi}{\pi}\frac{\sqrt{\sin\varphi}}{(E\cos^{2}\varphi+\sin\varphi)^{3/2}}\stackrel{E\ll1}{\longrightarrow}\frac{2}{\pi}\ln\!\left(\frac{c}{E}\right),\label{L2}\\
L_{3}(E) & =\int_{0}^{\pi}\frac{d\varphi}{\pi}\frac{(\sin\varphi)^{3/2}}{(E\cos^{2}\varphi+\sin\varphi)^{5/2}}\stackrel{E\ll1}{\longrightarrow}\frac{2}{\pi}\ln\!\left(\frac{d}{E}\right),
\end{align}
with the numerical constants $c=1.1$, $d=0.56$, and $F\equiv\ln{[(d^{3}\varepsilon_{0}^{2})/(\Lambda^{2}c)]}.$

Finally, the second term in Eq.~(\ref{Sigma3}) gives an induced
mass in the $p_{y}$ direction, which is generated by interactions,
\begin{align}
\Sigma_{y,m} & =-\frac{\alpha}{8}\int_{E_{\omega}}^{E_{\Lambda}}\frac{dE}{E}L_{2}(E)+\frac{3\alpha}{8}\int_{E_{\omega}}^{E_{\Lambda}}\frac{dE}{E}L_{4}(E)\nonumber \\
 & =-\frac{\alpha}{8\pi}\ln^{2}{(\Lambda/\omega)}-\frac{\alpha}{4\pi}G\ln{(\Lambda/\omega)},\label{Sigmaym}
\end{align}
where
\begin{equation}
L_{4}(E)=\int_{0}^{\pi}\frac{d\varphi}{\pi}\frac{E\sqrt{\sin\varphi}\cos^{2}\varphi}{(E\cos^{2}\varphi+\sin\varphi)^{5/2}}\stackrel{E\ll1}{\longrightarrow}\frac{4}{3\pi},\label{L4}
\end{equation}
and $G\equiv\ln{[c\varepsilon_{0}/\Lambda]}-2$. 

On the basis of the above results, the renormalized  Hamiltonian
to leading order in the interaction has the form: 
\begin{equation}
\mathcal{H}(\mathbf{p})=\left(g_{1}(\omega)\frac{p_{x}^{2}}{2}-g_{2}(\omega)\frac{p_{y}^{2}}{2}\right)\hat{\sigma}_{x}+v(\omega)p_{y}\hat{\sigma}_{y}.\label{ham-rg}
\end{equation}
We define the inverse masses as 
\begin{equation}
g_{1}(\omega)=m_{x}^{-1}(\omega),\ \ g_{2}(\omega)=m_{y}^{-1}(\omega).
\end{equation}
The functions $g_{1}(\omega), g_{2}(\omega), v(\omega)$ will be found below from the  solution
of the RG equations. 
The bare values of all parameters, i.e. the values at the lattice
(ultraviolet) energy scale, are determined by the parameters of the
Hamiltonian without interactions: $g_{1}(\omega=\Lambda)\equiv g_{10}=m^{-1}$,
$g_{2}(\omega=\Lambda)\equiv g_{20}=0$ and $v(\omega=\Lambda)\equiv v_{0}=v$.
Similarly,  $\alpha_{0} = e^2/v = \alpha$ (from Eq.~(\ref{coupling})) is defined as the bare
 value of the Coulomb coupling, corresponding to the bare value of $v_0=v$.

Taking into account  Eqs.~(\ref{Sigma3})$-$(\ref{L4}), we
have the one-loop perturbation theory results:
\begin{equation}
v(\omega)=v\left(1+\frac{\alpha}{\pi}\ln(\Lambda/\omega)\right),\label{v-leading}
\end{equation}
\noindent 
\begin{eqnarray}
g_{1}(\omega) & = & g_{10}\!\left(1+\frac{\alpha}{4\pi}\ln^{2}(\Lambda/\omega)+\frac{\alpha}{4\pi}F\ln(\Lambda/\omega)\right),\qquad\label{M-1}
\end{eqnarray}
and
\begin{eqnarray}
g_{2}(\omega) & = & g_{10}\!\left(\frac{\alpha}{8\pi}\ln^{2}(\Lambda/\omega)+\frac{\alpha}{4\pi}G\ln(\Lambda/\omega)\right).\qquad\label{my}
\end{eqnarray}
Here $g_{10}=m^{-1}$, as explained previously. 
 We find, in particular,  that a mass
term is generated in the $p_{y}$ direction, where the dispersion
was originally linear. The most important feature of the mass renormalization
formulas above is that both masses contain a log squared contribution
at leading order in the coupling $\alpha$. In addition, the two mass
terms have different signs upon renormalization (with the sign 
in front of $m_{y}$ being negative). In Eqs.~(\ref{M-1},\ref{my})
we have also kept sub-leading (first power) log contributions which 
strictly speaking is not necessary; however we retain them in our
calculations for completeness.


\begin{figure*}
\begin{centering}
\includegraphics[scale=0.53,angle=90]{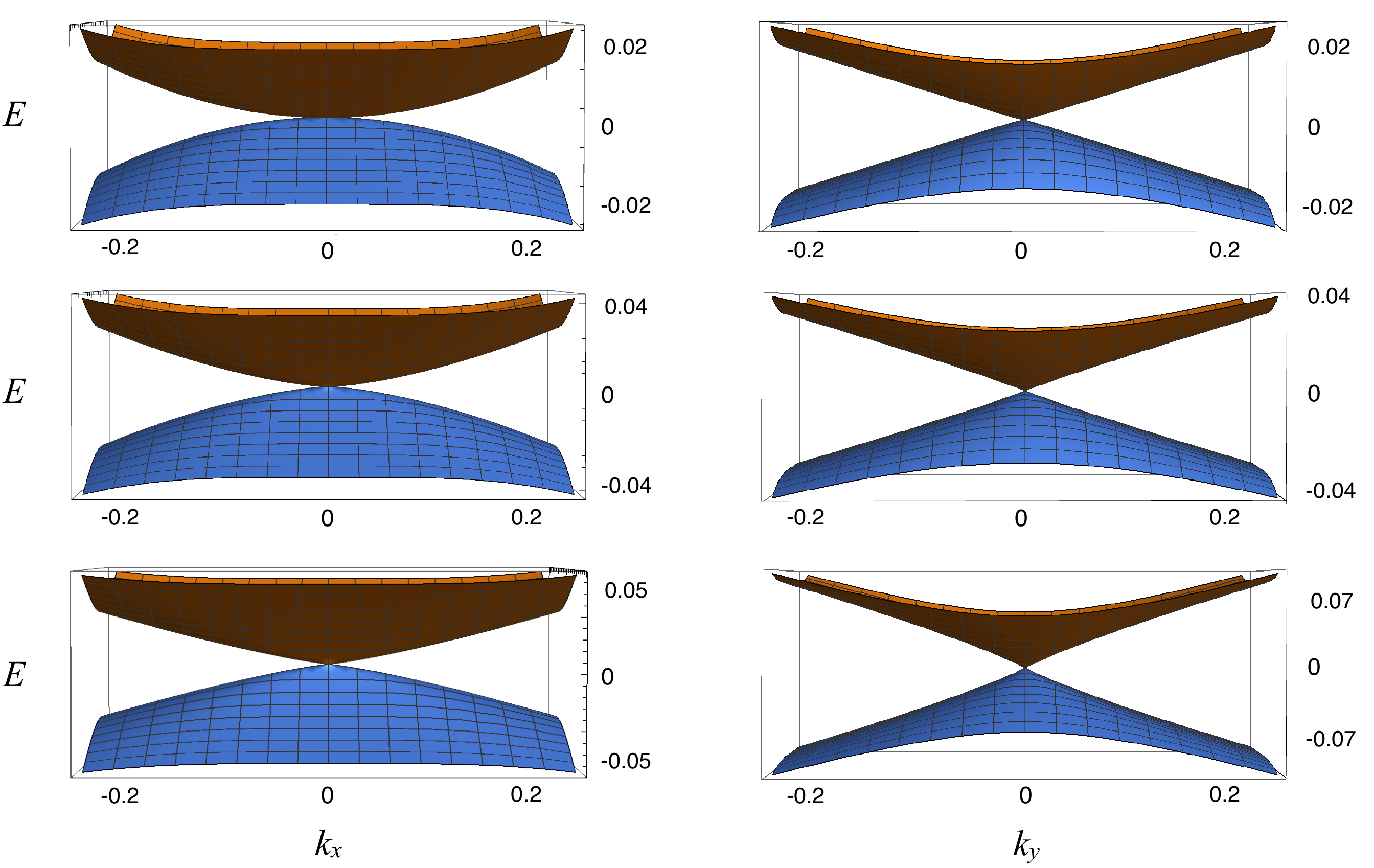}
\par\end{centering}
\caption{\label{fig:RG-Spectrum} Evolution of the renormalized electronic
spectrum $E=\tilde{\varepsilon}({\bf k})$ in the $k_{x}$ and $k_{y}$ directions. Energy is 
in units of $\Lambda/2$ and momenta are in units of inverse lattice spacing; the values of
various parameters ($\Lambda, m, v$) are fixed as described in the text. Top row: non-interacting
case, $\alpha/\pi=0$. Mid row: $\alpha/\pi=0.16$. Bottom row: $\alpha/\pi=0.63$.
The left column shows the transition from parabolic to linear dispersion,
driven by the mass renormalization, as the interaction increases.
The right column indicates the logarithmic velocity renormalization,
as in graphene.}

\end{figure*}


\subsection{Renormalization Group Equations and their Solutions}
\label{rg-results}

Given that the Coulomb interaction in 2D is a non-analytic function,
the electron charge does not renormalize \cite{Herbut2006,Ye1998} in the RG flow. Next, define the RG scale
\begin{equation}
\ell\equiv\ln{(\Lambda/\omega)}.\label{rg-scale}
\end{equation}
From Eqs.~(\ref{v-leading})$-$(\ref{my}), we obtain the RG equations
\begin{equation}
\frac{dv(\ell)}{d\ell}=v(\ell)\alpha(\ell)/\pi=e^{2}/\pi,\label{v-rg}
\end{equation}
\begin{equation}
\frac{dg_{1}(\ell)}{d\ell}=g_{1}(\ell)\left(\frac{\alpha(\ell)}{2\pi}\ell+\frac{\alpha(\ell)}{4\pi}F\right),\label{g1-rg}
\end{equation}
and 
\begin{equation}
\frac{dg_{2}(\ell)}{d\ell}=g_{1}(\ell)\left(\frac{\alpha(\ell)}{4\pi}\ell+\frac{\alpha(\ell)}{4\pi}G\right).\label{g2-rg}
\end{equation}
Integrating the velocity, Eq.~(\ref{v-rg}), we obtain 
\begin{equation}
v(\ell)=v\left(1+\frac{\alpha}{\pi}\ell\right)\ \Rightarrow\ \alpha(\ell)=\frac{\alpha}{1+\frac{\alpha}{\pi}\ell},
\end{equation}
which in turn determines the running of the interaction coupling constant. 
Similarly to isotropic graphene, the velocity increases logarithmically
as energy decreases, leading to a logarithmic decrease of the interaction,
which flows to weak coupling,
\begin{equation}
\alpha(\ell=\ln{(\Lambda/\omega)})=\frac{\alpha}{1+\frac{\alpha}{\pi}\ln{(\Lambda/\omega)}}.
\end{equation}

Eq.~(\ref{g1-rg}) can be integrated with the result
\begin{equation}
g_{1}(\ell)/g_{10}=\left(1+\frac{\alpha}{\pi}\ell\right)^{F/4}e^{\ell/2-\frac{\pi}{2\alpha}\ln{\left(1+\frac{\alpha}{\pi}\ell\right)}}.
\end{equation}
Rewriting this result as a function of energy $\omega$, by taking
into account Eq.~(\ref{rg-scale}), we obtain: 
\begin{equation}
g_{1}(\omega)/g_{10}=\frac{\sqrt{\Lambda/\omega}}{\left(1+\frac{\alpha}{\pi}\ln{(\Lambda/\omega)}\right)^{(\pi/2\alpha)-F/4}}.\label{g1-rg-solution}
\end{equation}
It is instructive to expand Eq.~(\ref{g1-rg-solution}) for small
values of the bare coupling (we set $F=0$ in this formula for clarity),
\begin{eqnarray}
 &  & g_{1}(\omega)/g_{10}\approx1+\frac{\alpha}{4\pi}\ln^{2}{(\Lambda/\omega)}+\nonumber \\
 &  & +\frac{\alpha^{2}}{\pi^{2}}\left(\frac{1}{32}\ln^{4}{(\Lambda/\omega)}-\frac{1}{6}\ln^{3}{(\Lambda/\omega)}\right)+O(\alpha^{3}),
\end{eqnarray}
which gives an idea of the structure of higher orders of perturbation
theory, re-summed by the RG. We note that the expansion is well controlled
at all orders when $\alpha/\pi\ll1$. This inequality defines
the validity of the perturbative regime.

The RG solution, Eq. (\ref{g1-rg-solution}), is one of our main results.
Examining the $p_{x}$ direction part of the dispersion, we clearly
see that in the low energy limit $\Lambda/\omega\gg1$, we have the
dominant behavior $g_{1}(\omega)/g_{10}\sim\sqrt{\Lambda/\omega}$,
up to logarithmic corrections. This in turn implies that the mass
term in the renormalized Hamiltonian (\ref{ham-rg}), which has the
structure $g_{1}(\omega)\frac{p_{x}^{2}}{2}$, effectively becomes
linear in momentum when the energy is on-shell, as defined in Eq.~(\ref{eq:omega}).
More precisely, for low momenta, $|p_{x}|/\sqrt{2m\Lambda}\ll1$, provided
also $\frac{\alpha}{\pi}\ln{(2m\Lambda/p_{x}^{2})}\gg1$, we have
\begin{equation}
g_{1}(\omega)\frac{p_{x}^{2}}{2}=\sqrt{\frac{\Lambda}{2m}}\frac{|p_{x}|}{\left[\frac{\alpha}{\pi}\ln{(2m\Lambda/p_{x}^{2})}\right]^{(\pi/2\alpha)-F/4}}.
\end{equation}
We see from here that the dispersion becomes linear, with log correction
whose power depends on the value of $\alpha$ (the value of the
subleading piece $F$ is conceptually and numerically not important;
for our parameter values we have $|F|/4\approx0.1$). Therefore for
small values of $\alpha$ when the power $\frac{\pi}{2\alpha}$
is large, the log term presence will provide some bending to the dispersion;
as $\alpha$ increases the linearity becomes gradually more pronounced.
We will see shortly that the numerical plot of the RG dispersion confirms
this behavior.

Finally, integration of Eq.~(\ref{g2-rg}) 
\begin{equation}
g_{2}(\ell)=\frac{1}{4\pi}\int_{0}^{\ell}[\xi g_{1}(\xi)\alpha(\xi)+Gg_{1}(\xi)\alpha(\xi)]d\xi
\end{equation}
leads to a cumbersome expression which is not particularly illuminating
and will be taken into account numerically. We can deduce however
both analytically and numerically that in the extreme low energy limit
($\ell\rightarrow\infty$) 
\begin{equation}
g_{2}(\omega)/g_{1}(\omega)\rightarrow1/2,\ \ \omega\rightarrow0.
\end{equation}
This is related to the factor of two difference which appears in the
RG Equations~(\ref{g1-rg},\ref{g2-rg}). Therefore the induced $g_{2}$
term plays a marginal role and modifies somewhat the dispersion in
the $p_{y}$ direction at intermediate energies, while at low energies
it does not change the preexistent linear behavior.

Our numerical results for the renormalized dispersion, 
\begin{equation}
\tilde{\varepsilon}({\bf p})=\pm\sqrt{\left (\frac{g_{1}(\omega)p_{x}^{2}}{2}-\frac{g_{2}(\omega)p_{y}^{2}}{2} \right )^{2}+v(\omega)^{2}p_{y}^{2}},\label{epsilon}
\end{equation}
evaluated simultaneously with Eq.~(\ref{eq:omega}), are presented
in Fig.~\ref{fig:RG-Spectrum}. We use the following values of parameters
for these plots, setting $v=1$: $m=2,\Lambda=2,F=-0.4,G=-1.2$. In
the full units, $mv=2\hbar/a$, $\Lambda=2\hbar v/a$, where $a$
is the lattice spacing. The overall behavior is quite robust and not
sensitive to these particular values (in particular the subleading
pieces $F,G$ follow from the previously derived formulas and are
non-universal, although the results are very weakly dependent on their
exact values, as expected). We see that the spectrum undergoes a profound
transformation from parabolic towards linear, thus recovering a more
conventional Dirac cone shape. In the $p_{y}$ direction the spectrum
remains linear even though it undergoes renormalization due to the
increase of the velocity at low energy.

\begin{figure}
\includegraphics[width=0.49\textwidth]{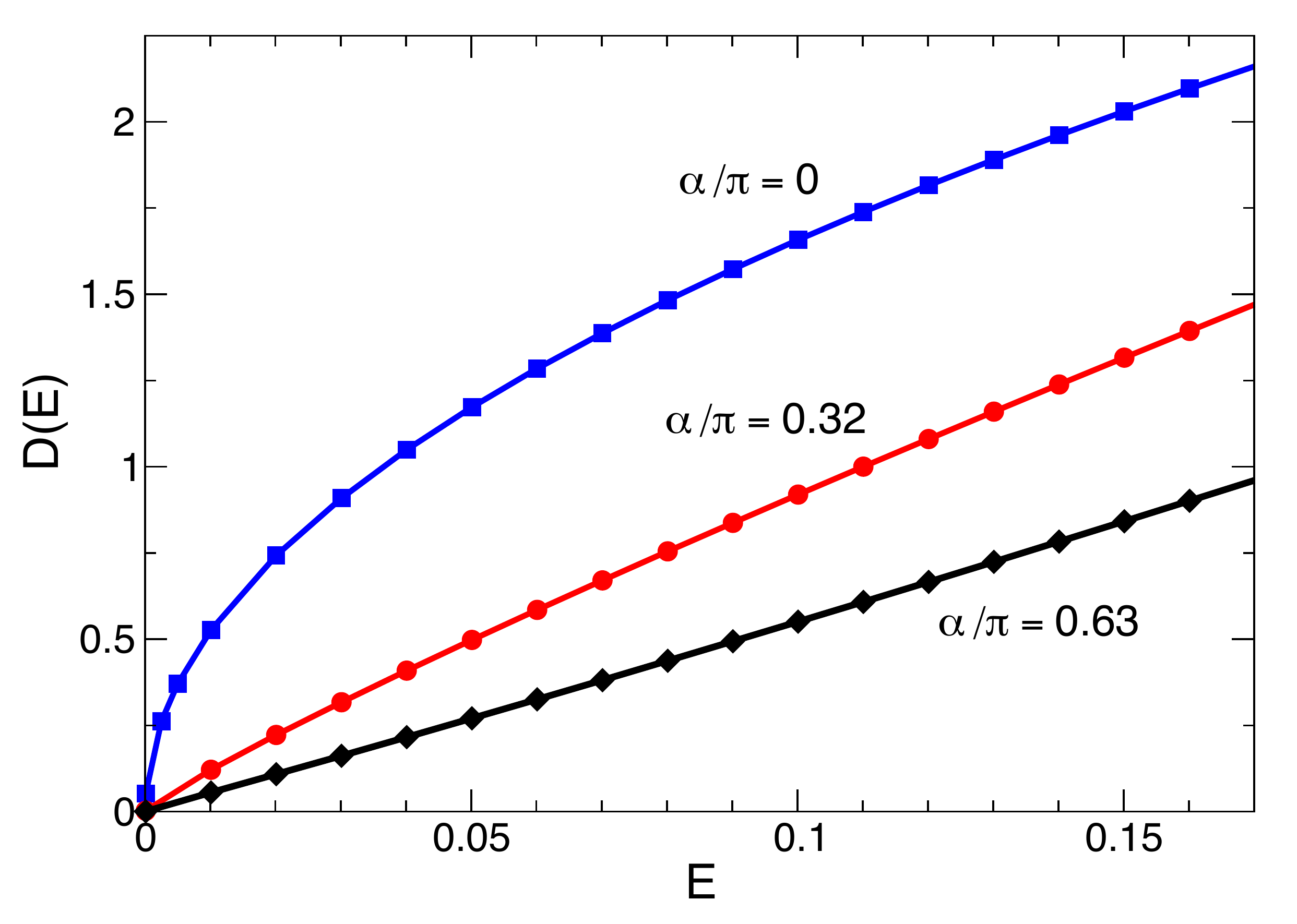} 
\caption{\label{fig:RG-DOS}Renormalized density of states as a function of
energy $D(E)$, where  the energy $E$ is 
in units of $\Lambda$ and $D(E)$ is in units of $\sqrt{2m\Lambda}/[(2\pi)^2v]$.
We show the transition from $D(E)\propto\sqrt{E}$ in the
absence of interactions ($\alpha/\pi=0$) to linear behavior $D(E)\propto E$
$(\alpha/\pi=0.63$) as the interaction coupling increases.}
\end{figure}

A different way to detect the transition towards Dirac cone behavior
is to monitor the density of states (DOS) which can be expressed in the
following way for the renormalized spectrum: 
\begin{equation}
D(E)=\frac{\sqrt{2m}}{v(2\pi)^2}\int d\varepsilon\int_{0}^{\pi}d\varphi\frac{\sqrt{\varepsilon}}{\sqrt{\sin\varphi}}\delta(E-\tilde{\varepsilon}(\varepsilon,\varphi)).
\end{equation}
Here the notation $\tilde{\varepsilon}(\varepsilon,\varphi)$ means
that the momenta are expressed via the energy-angle variables as in
Eqs.~(\ref{var-change},\ref{var-change-int}). Without interactions
($\alpha=0$) we have $\tilde{\varepsilon}(\varepsilon,\varphi)=\varepsilon$
by the very definition of the energy-angle variables and we obtain
the well-known result for a semi-Dirac dispersion, $D(E)\sim\sqrt{E}$.
As the interaction $\alpha$ increases we evaluate the above formula
numerically and see quite clearly the transition to linear behavior,
as shown in Fig.~\ref{fig:RG-DOS}.


 Finally, we calculate the Berry phase associated with the renormalized Hamiltonian. 
As is well known, the Berry phase  is given by  the circulation of the  wave-function phase gradient around
the Fermi point ($k_x=k_y=0$), or more explicitly 
 $\frac{1}{2} \oint (\bm{\nabla} \phi({\bf{k}})).{\mbox{d}} \bf k$. The Hamiltonian (both bare and renormalized)
 has the form $\mathcal{H}(\mathbf{k})=h_x(\mathbf{k}) \hat{\sigma}_{x}+ h_y(\mathbf{k}) \hat{\sigma}_{y}$. Then the phase of the wave function is determined by the equation:  
$\tan{\phi({\bf{k}})}=h_y(\mathbf{k})/h_x(\mathbf{k})$. Consequently one finds that the Berry phase is zero both 
for the bare and renormalized semi-Dirac cases. For the bare case it was understood  a while back \cite{Montambaux2009}
that since the semi-Dirac spectrum appears as a merger of two Dirac cones with Berry phases $\pm \pi$ (related by time-reversal  symmetry), at the topological Lifshitz point the Berry phase is zero, being a sum of those two values. Technically
this is related to the fact that $h_x(\mathbf{k})$ is even under the transformation $k_x \rightarrow -k_x$ for semi-Dirac fermions, leading to zero  Berry phase. Even though the spectrum undergoes complex renormalization when Coulomb
interactions are included, the above parity symmetry is preserved in the renormalized Hamiltonian and we find that the Berry phase is identically zero.  This is natural since the Berry phase is a purely topological property and should not change
upon introduction of (parity and time-reversal preserving) interaction effects.

\section{Implications for screening and self-consistency}
\label{screening}

Let us also discuss more precisely the region of applicability of our results. We use perturbation theory to leading
order with the bare Coulomb interaction and it is therefore important to assess the effect of screening.  
We have calculated the static polarization function $\Pi({\bf q})$ numerically and  found that it 
has the expected form \begin{equation}
\Pi({\bf p}) =  - N \frac{C}{4} \frac{\sqrt{2m}}{v}\sqrt{|\varepsilon({\bf p})|},
\end{equation}
consistent with the scaling of the density of states. These results are also in agreement with the literature 
\cite{Cho2016,Wang17}. In this formula $C$ has a very weak dependence on the direction in ${\bf p}$ space,
deviating slightly from the value $C\approx 0.25$.
The screened potential within the random phase approximation (RPA)  becomes ($\alpha = e^2/v$):
\begin{equation}
V_{RPA}({\bf p}) \!= \!\frac{2\pi e^2}{|{\bf p}| -2\pi e^2 \Pi({\bf p}) } =  \frac{2\pi e^2}{|{\bf p}| + \frac{C \pi \sqrt{2m}}{2} (N\alpha)
 \sqrt{|\varepsilon({\bf p})|}}
\end{equation}
 Therefore in the $p_x$ direction (setting $p_y=0$ in the above formula), we see that screening is purely dielectric (momentum independent). The condition that
 the bare term dominates over the polarization, i.e. $|{\bf p}| \gg \frac{C \pi \sqrt{2m}}{2} (N\alpha) \sqrt{|\varepsilon({\bf p})|}$, translates into
 the condition $C (\pi/2) N\alpha \ll 1$ which is the starting point of our calculation.  
 On the other hand in the $p_y$ direction screening is present, and the bare term is dominant provided $p_y \gg  [C (\pi/2)\sqrt{2mv}(N\alpha)]^2 \equiv p_{y, min}$, 
 which defines  the momentum $p_{y, min}$. 
 
 Below this small  momentum scale,  $p_{y, min} \sim (N\alpha)^2 \ll 1$, it is tempting to conclude that bare perturbation theory is invalid. 
 The bare perturbative analysis of the polarization bubble, however, is incomplete. In the spirit of RG, one must account for the self-consistent renormalization of all physical observables, reflecting an exact resummation of leading logarithmic divergences in all orders of perturbation theory. In that philosophy, one must account for the effects of the velocity and mass renormalization in the polarization bubble, and consider it explicitly in the analysis of any screening effects in the RG results. 
 
 We found in Section \ref{rg-results}  that the spectrum
  undergoes very strong renormalization at low energy, with the linear dispersion effectively restored (Figs.~\ref{fig:RG-Spectrum},\ref{fig:RG-DOS}). 
  Therefore  a ``renormalized" RPA potential $\tilde{V}_{RPA}({\bf p})$ has to be constructed 
  based on the renormalized  $\tilde{\Pi}({\bf p})$, which could change significantly the  structure of the bare RPA potential.
   Qualitatively we expect the following behavior:
  since the density of states undergoes a crossover to  linear behavior (Fig.~\ref{fig:RG-DOS})
   at finite coupling $\alpha/\pi$ (which follows the crossover in the spectrum itself, Fig.~\ref{fig:RG-Spectrum}),
 then we expect  
 \begin{equation}\tilde{\Pi}({\bf p}) \sim -\frac{N}{v_x(\alpha)v_y(\alpha)}\sqrt{[v_x(\alpha)]^2p_x^2+[v_y(\alpha)]^2p_y^2}. 
 \end{equation} 
 This formula  reflects the fact that the renormalized dispersion is characterized by effective
 (possibly coupling-dependent) velocities $v_x(\alpha), v_y(\alpha)$ in both directions and therefore the polarization would have the well-known functional form  for anisotropic 
 Dirac fermions.   Consequently, 
 \begin{equation}
\tilde{V}_{RPA}({\bf p}) \! \!=\!\! \frac{2\pi e^2}{|{\bf p}| + \frac{\tilde{C}(\alpha)v}{v_x(\alpha)v_y(\alpha)} (N\alpha)
 \sqrt{[v_x(\alpha)]^2p_x^2+[v_y(\alpha)]^2p_y^2}}.
\end{equation}
This formula is valid  at finite $\alpha$ only, reflecting the dressed (beyond RPA) polarization structure. $\tilde{C}(\alpha)$ is a function that could also show
some weak angular dependence and is not important for our intuitive argument. In addition, it is known that the anisotropy in the Dirac spectrum tends to 
disappear under renormalization \cite{Vafek2002} (i.e. $v_x/v_y \rightarrow 1$).  

From these considerations, we conclude that one can expect simple dielectric screening 
 at weak coupling. Hence, our analysis leads to a fully self-consistent picture, i.e.  is valid all the way down to zero energy, where screening (calculated self-consistently) is not important. 
 Therefore the low-energy RG equations discussed in Section \ref{rg-results} represents the true RG fixed point behavior in the perturbative regime of the problem.


\begin{figure}
\includegraphics[width=0.47\textwidth]{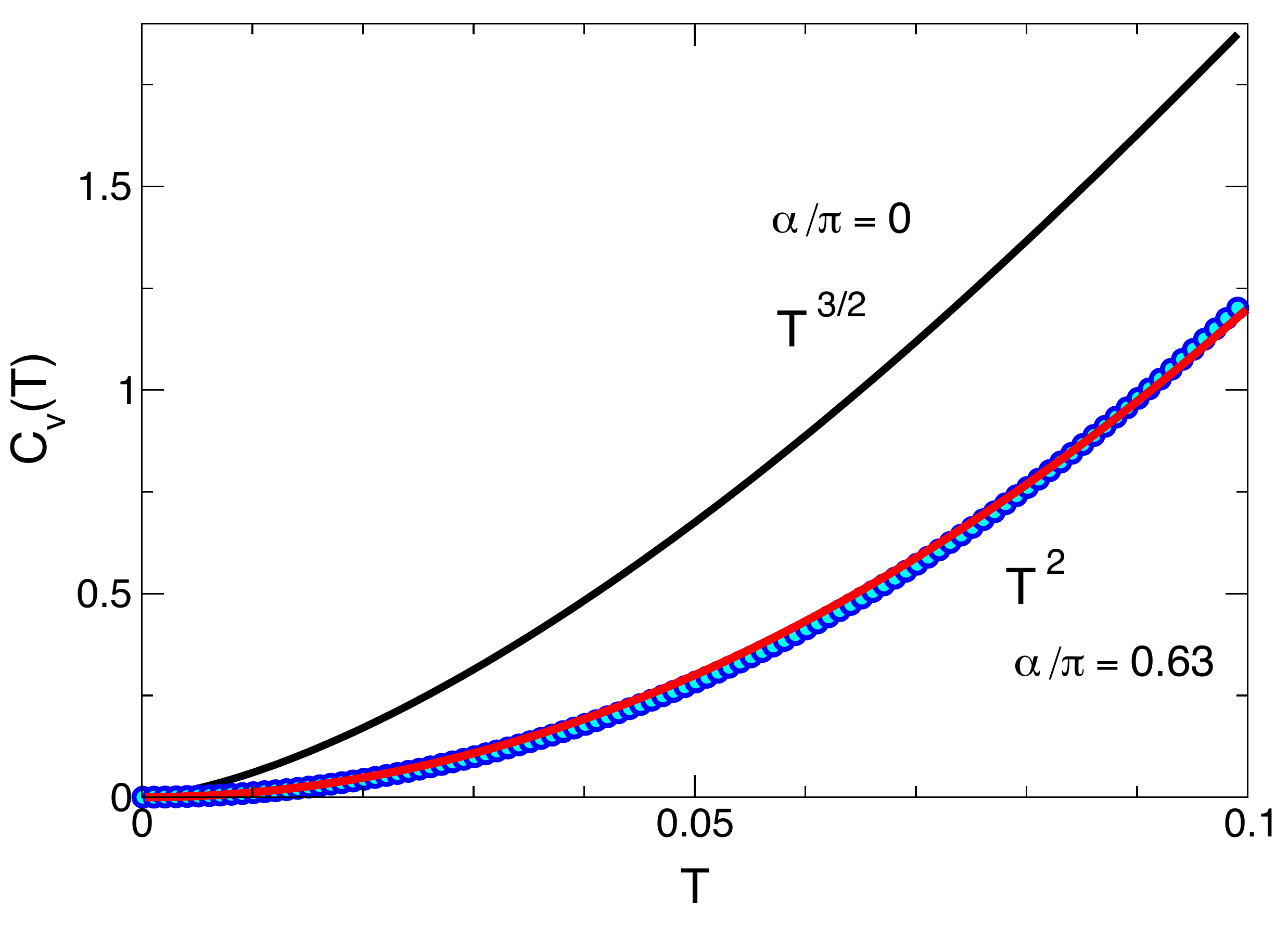}
\includegraphics[width=0.47\textwidth]{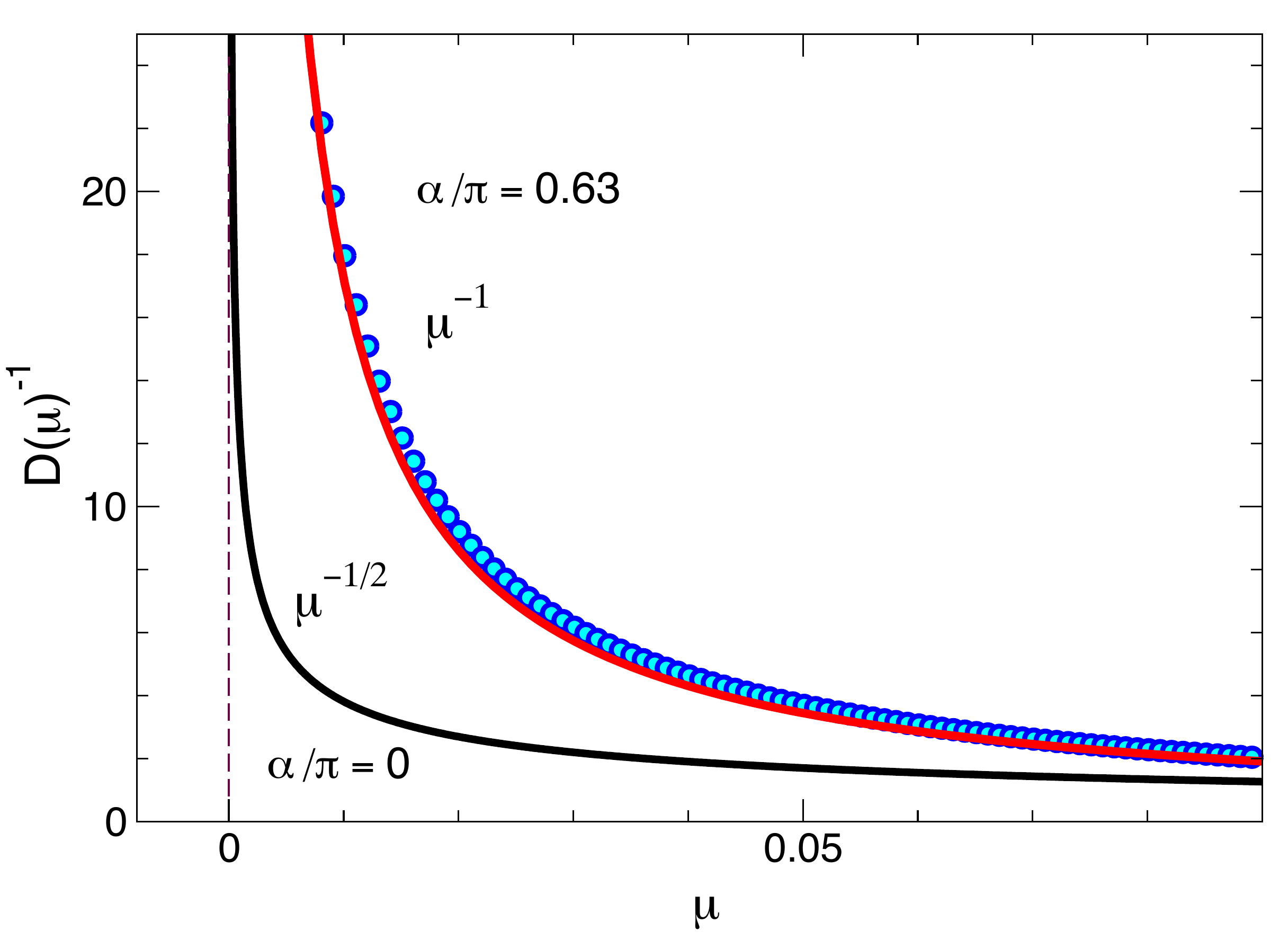}
\caption{\label{fig:observables}
Upper panel: Specific heat $C_V(T)$, in units of $\sqrt{2m}\Lambda^{3/2}/[(2\pi)^2v]$,  evaluated for non-interacting semi-Dirac fermions
($T^{3/2}$ law),  and for finite value of the interaction, leading to behavior ($T^2$) consistent with linear Dirac dispersion. Symbols represent 
numerical evaluation and the solid red line is the pure $T^2$ behavior. Temperature is measured in units of $\Lambda$.
Lower panel: Inverse DOS, $\partial \mu/\partial n = D(\mu)^{-1}$, as a function of the chemical potential showing the non-interacting 
behavior ($\mu^{-1/2}$, shifted by a factor of two for clarity),  changing to $\mu^{-1}$ (characteristic of linear Dirac fermions)  upon renormalization. Symbols represents
numerical evaluation and the solid red line is the pure $\mu^{-1}$ behavior. DOS and energy units are the same as in Fig.~\ref{fig:RG-DOS}.
}
\end{figure}


\section{Physical Observables}
\label{observables}

Here we discuss the effect of the strong spectrum renormalization on physical observables and potential relevance to real materials.
The specific heat low temperature dependence $C_V(T)$ is sensitive to the low energy dispersion. It can be computed via $C_V(T)=-T\partial^2 F/\partial T^2$, where
$F$ is the free energy. 
One then obtains the standard formula for fermionic quasiparticles, 
\begin{equation} C_V(T) \sim T^{-2} \int (d^2k/4\pi^2) \varepsilon({\bf k})^2\cosh^{-2}{(\varepsilon({\bf k})/2T)},
\end{equation}
which leads to the following results for  semi-Dirac fermions before and after renormalization (upon replacing the bare with the renormalized dispersion, $\varepsilon({\bf k}) \rightarrow \tilde{\varepsilon}({\bf k}))$:
\begin{equation}
C_V(T) \sim T^{3/2},  \  \  \ {\mbox{bare semi-Dirac}}
\end{equation}
\begin{equation}
C_V(T) \sim T^2,  \  \  \ {\mbox{renormalized.}}
\end{equation}
The last formula reflects the crossover towards linear behavior  in the density of states at finite $\alpha$ upon renormalization (Fig.~\ref{fig:RG-DOS}) and represents
the result for linear Dirac fermions. 
Fig. \ref{fig:observables} shows this behavior in more detail, comparing the numerical evaluation of $C_V$ with the renormalized dispersion and the pure $T^2$ law, similar
to graphene.

The  electronic compressibility $\kappa$, measured for example by quantum capacitance techniques, is also very sensitive to the dispersion 
and interaction effects in general  \cite{Yacoby08,Geim13,Sheehy07,Kotov2012}.
It is  defined as $\kappa^{-1}=n^2 (\partial \mu/\partial n)$, however the charge response is experimentally determined by 
$\partial \mu/\partial n$, which is the inverse density of states, related to the inverse capacitance as explained in the above literature. Therefore for the charge response we obtain
\begin{equation}
  \frac{\partial \mu}{\partial n} \sim \frac{1}{\sqrt{\mu}} \sim \frac{1}{n^{1/3}},  \  \  \ {\mbox{bare semi-Dirac}}
\end{equation}
\begin{equation}
\frac{\partial \mu}{\partial n}  \sim \frac{1}{\mu} \sim  \frac{1}{\sqrt{n}},  \  \  \ {\mbox{renormalized}}.
\end{equation}
Fig. \ref{fig:observables} shows the behavior of the inverse DOS, $\partial \mu/\partial n$, 
as a function of the chemical potential $\mu$ relative to the Dirac point, at zero temperature. Clearly
the behavior associated with linear dispersion, $D(\mu) \sim \mu$, is observed for interacting
 renormalized fermions. Finally, it is often useful experimentally to plot $\partial \mu/\partial n$
 as a function of the electron density $n$. The relevant dependence is also shown in the above equations
 where we have used the relationship between the chemical potential and density: 
 for semi-Dirac fermions $\mu \sim n^{2/3}$, and for linear Dirac fermions $\mu \sim \sqrt{n}$.
 Our renormalized theory clearly 
predicts power laws similar to graphene \cite{Yacoby08,Geim13,Sheehy07}.
The above formulas can also be written as a function of temperature $T$ at $\mu=0$, where we 
have the corresponding behavior: $\partial \mu/\partial n \sim 1/\sqrt{T}$ for the  bare semi-Dirac dispersion
 and $ \sim 1/T$  for our renormalized case.

Thus we conclude that physical observables  associated with interacting fermions show the characteristic power laws associated with linear Dirac dispersion at low energy and therefore they can be clearly distinguished from the
different powers in the case of non-interacting semi-Dirac fermions. Our results are also  very different  from the large-$N$ theory \cite{Isobe2016,Cho2016} which predicts 
power law behavior similar to the  non-interacting semi-Dirac case,  with powers modified by  small  corrections  of order $1/N$.

In real materials such as black phosphorus under doping \cite{Kim723}, the Fermi velocity has been measured by angle resolved photoemission spectroscopy  (ARPES) to be  $v\approx 5 \times 10^5$ m/s over an energy window of $\sim$1eV around the touching point of the bands.  This value is approximately half of the one measured  in graphene and  corresponds to  an effective fine structure constant $\alpha/\pi \approx  1.4 /\epsilon$, where $\epsilon$ is the dielectric constant due to  screening effects. We expect that relatively weak dielectric screening could lead to values of $\alpha/\pi$  that fall within the range where perturbation theory is valid. We note that the restoration of the linearity in the spectrum is detectable within a much narrower energy window around the neutrality point compared to the typical energy window investigated with ARPES. We propose that quantum capacitance measurements  \cite{Yacoby08,Geim13} of the electronic compressibility would have enough energy resolution to reveal the low energy behavior of the electronic dispersion in the perturbative regime.

\section{Gap Renormalization away from criticality}

For completeness, we also consider behavior away from the critical
point in order to assess how the gap changes under interaction-induced
renormalization. In fact we consider modification of the Hamiltonian
to include two gap-producing pieces: (1) $\Delta_{1}$, already mentioned
previously, and (2) $\Delta_{2}$, which could be generated by excitonic
pairing, 
\begin{equation}
\mathcal{H}(\mathbf{p})=\left(\frac{p_{x}^{2}}{2m}+\Delta_{1}\right)\hat{\sigma}_{x}+vp_{y}\hat{\sigma}_{y}+\Delta_{2}\hat{\sigma}_{z}.
\end{equation}
The spectrum now obviously becomes: 
\begin{equation}
\varepsilon({\bf p})=\sqrt{\left(\frac{p_{x}^{2}}{2m}+\Delta_{1}\right)^{2}+v^{2}p_{y}^{2}+\Delta_{2}^{2}}.\label{e1}
\end{equation}
The renormalization of the two gaps in the frequency regime of interest
\begin{equation}
\sqrt{\Delta_{1}^{2}+\Delta_{2}^{2}}\leq\omega\ll\Lambda,
\end{equation}
can be determined similarly to the procedure from the previous section.
We will  keep only the leading log contributions. Our final result
is 
\begin{equation}
\Delta_{1}(\omega)=\Delta_{1}\left(1+\frac{\alpha}{4\pi}\ln^{2}{(\Lambda/\omega)}+\cdots\right),
\end{equation}
\begin{equation}
\Delta_{2}(\omega)=\Delta_{2}\left(1+\frac{\alpha}{4\pi}\ln^{2}{(\Lambda/\omega)}+\cdots\right).
\end{equation}
This shows that the two gaps are renormalized exactly the same way
and again the unconventional log squared behavior is the dominant
one even at first order in the interaction. The next steps are identical
to the ones preformed in the previous section for the mass terms.
The corresponding RG equations are 
\begin{equation}
\frac{d\Delta_{i}(\ell)}{d\ell}=\Delta_{i}(\ell)\frac{\alpha(\ell)}{2\pi}\ell,\qquad(i=1,2).\label{DeltaRG}
\end{equation}
Their solution leads to the following results: 
\begin{equation}
\Delta_{i}(\omega)=\Delta_{i}\frac{\sqrt{\Lambda/\omega}}{\left(1+\frac{\alpha}{\pi}\ln{(\Lambda/\omega)}\right)^{(\pi/2\alpha)}},\quad(i=1,2).
\end{equation}
These demonstrate that if the initial ``bare" gaps ($\Delta_{1,2}$)
are present, the gap values will increase quite strongly $\sim\sqrt{\Lambda/\omega}$
under renormalization at low energy (with additional, interaction-dependent
log variation). In particular if $\Delta_{2}=0$ (no excitonic pairing),
the sign and value of $\Delta_{1}=\Delta$ controls the distance from
criticality ($\Delta<0$, gapless phase; $\Delta>0$, gapped phase)
and therefore if the system is initially on either side of criticality,
it will keep flowing away from it. Similarly, if excitonic pairing
is present, it will increase under renormalization. Such tendency
(for excitonic pairing) is similar to the case of graphene \cite{Kotov2012},
except that in our case the renormalization is much stronger (related
to the log squared behavior in perturbation theory).

\section{Conclusions}

We have performed a full RG analysis for semi-Dirac fermions at first
non-trivial order in the interaction. Our calculation is perturbative
($\alpha/\pi\ll1$), and it should be reliable for reasonably small
bare values of $\alpha/\pi$. The system subsequently flows towards
weak coupling under RG. The unconventional log squared behavior present
in the mass terms in bare perturbation theory translates into strong
(square root of energy scaling) mass renormalization in the full solution
of the RG equations (Eq.~(\ref{g1-rg-solution})). This behavior
effectively wipes out the curvature of the dispersion and the system
flows towards a Dirac dispersion which is anisotropic but linear in
momentum. However an additional logarithmic scaling with interaction-dependent
power exists on top of the linear momentum dispersion; as the interaction
increases the logarithmic part becomes less pronounced. Away from
the critical point in either direction, we find that gap renormalization
is also very strong and the system flows further away from criticality.

We have also presented arguments that our weak-coupling RG procedure is fully self-consistent
in a sense that if we dress the Coulomb potential with RPA corrections, it will eventually, upon renormalization,  become similar to the unscreened interaction. Therefore our low-energy RG behavior represents a true weak coupling fixed point.

The emergent, upon renormalization, linear Dirac fermions at the Lifshitz point are also
unusual in the sense that they carry zero Berry phase. This is a topological property that remains unaffected by our strong renormalization, since it is related to the fact that the original (non-interacting)
semi-Dirac fermions  arise from the merger of two Dirac cones, related by time-reversal, with opposite Berry phases.

Overall, we have shown that the full weak coupling RG implementation
gives results that are very different from the large $N$ approach,
which favors a fixed point with renormalized semi-Dirac dispersion and also exhibits incoherent (``NFL-MFL")
behavior.  Our results therefore can have profound consequences for
understanding systems with interacting semi-Dirac fermions. In particular we make clear predictions for physical observables, such as the  specific heat and electronic compressibility, which display characteristic power laws as a function of temperature or Fermi energy, consistent with linear Dirac dispersion.

\section*{acknowledgments}

V.N.K. gratefully acknowledges the financial support of the Gordon Godfrey
visitors program at the School of Physics, University of New South
Wales, Sydney, during two research visits. V.N.K. also acknowledges partial financial support from NASA Grant No. 80NSSC19M0143 during the final stages of this work. B.U. acknowledges the Carl T.
Bush fellowship for partial support. B.U. also acknowledges NSF Grant No. DMR-2024864 for support. O.P.S. was supported by the Australian Research Council
Centre of Excellence in Future Low Energy Electronics
Technologies (Grant No. CE170100039).

\appendix
\bibliographystyle{apsrev4-1}
\bibliography{refs}

\begin{thebibliography}{36}%
\makeatletter
\providecommand \@ifxundefined [1]{%
 \@ifx{#1\undefined}
}%
\providecommand \@ifnum [1]{%
 \ifnum #1\expandafter \@firstoftwo
 \else \expandafter \@secondoftwo
 \fi
}%
\providecommand \@ifx [1]{%
 \ifx #1\expandafter \@firstoftwo
 \else \expandafter \@secondoftwo
 \fi
}%
\providecommand \natexlab [1]{#1}%
\providecommand \enquote  [1]{``#1''}%
\providecommand \bibnamefont  [1]{#1}%
\providecommand \bibfnamefont [1]{#1}%
\providecommand \citenamefont [1]{#1}%
\providecommand \href@noop [0]{\@secondoftwo}%
\providecommand \href [0]{\begingroup \@sanitize@url \@href}%
\providecommand \@href[1]{\@@startlink{#1}\@@href}%
\providecommand \@@href[1]{\endgroup#1\@@endlink}%
\providecommand \@sanitize@url [0]{\catcode `\\12\catcode `\$12\catcode
  `\&12\catcode `\#12\catcode `\^12\catcode `\_12\catcode `\%12\relax}%
\providecommand \@@startlink[1]{}%
\providecommand \@@endlink[0]{}%
\providecommand \url  [0]{\begingroup\@sanitize@url \@url }%
\providecommand \@url [1]{\endgroup\@href {#1}{\urlprefix }}%
\providecommand \urlprefix  [0]{URL }%
\providecommand \Eprint [0]{\href }%
\providecommand \doibase [0]{http://dx.doi.org/}%
\providecommand \selectlanguage [0]{\@gobble}%
\providecommand \bibinfo  [0]{\@secondoftwo}%
\providecommand \bibfield  [0]{\@secondoftwo}%
\providecommand \translation [1]{[#1]}%
\providecommand \BibitemOpen [0]{}%
\providecommand \bibitemStop [0]{}%
\providecommand \bibitemNoStop [0]{.\EOS\space}%
\providecommand \EOS [0]{\spacefactor3000\relax}%
\providecommand \BibitemShut  [1]{\csname bibitem#1\endcsname}%
\let\auto@bib@innerbib\@empty
\bibitem [{\citenamefont {Montambaux}\ \emph
  {et~al.}(2009{\natexlab{a}})\citenamefont {Montambaux}, \citenamefont
  {Pi{\'e}chon}, \citenamefont {Fuchs},\ and\ \citenamefont
  {Goerbig}}]{Montambaux2009}%
  \BibitemOpen
  \bibfield  {author} {\bibinfo {author} {\bibfnamefont {G.}~\bibnamefont
  {Montambaux}}, \bibinfo {author} {\bibfnamefont {F.}~\bibnamefont
  {Pi{\'e}chon}}, \bibinfo {author} {\bibfnamefont {J.-N.}\ \bibnamefont
  {Fuchs}}, \ and\ \bibinfo {author} {\bibfnamefont {O.~M.}\ \bibnamefont
  {Goerbig}},\ }\href {\doibase 10.1140/epjb/e2009-00383-0} {\bibfield
  {journal} {\bibinfo  {journal} {The European Physical Journal B}\ }\textbf
  {\bibinfo {volume} {72}},\ \bibinfo {pages} {509} (\bibinfo {year}
  {2009}{\natexlab{a}})}\BibitemShut {NoStop}%
\bibitem [{\citenamefont {Montambaux}\ \emph
  {et~al.}(2009{\natexlab{b}})\citenamefont {Montambaux}, \citenamefont
  {Pi\'echon}, \citenamefont {Fuchs},\ and\ \citenamefont
  {Goerbig}}]{Montambaux2009-2}%
  \BibitemOpen
  \bibfield  {author} {\bibinfo {author} {\bibfnamefont {G.}~\bibnamefont
  {Montambaux}}, \bibinfo {author} {\bibfnamefont {F.}~\bibnamefont
  {Pi\'echon}}, \bibinfo {author} {\bibfnamefont {J.-N.}\ \bibnamefont
  {Fuchs}}, \ and\ \bibinfo {author} {\bibfnamefont {M.~O.}\ \bibnamefont
  {Goerbig}},\ }\href {\doibase 10.1103/PhysRevB.80.153412} {\bibfield
  {journal} {\bibinfo  {journal} {Phys. Rev. B}\ }\textbf {\bibinfo {volume}
  {80}},\ \bibinfo {pages} {153412} (\bibinfo {year}
  {2009}{\natexlab{b}})}\BibitemShut {NoStop}%
\bibitem [{\citenamefont {Adroguer}\ \emph {et~al.}(2016)\citenamefont
  {Adroguer}, \citenamefont {Carpentier}, \citenamefont {Montambaux},\ and\
  \citenamefont {Orignac}}]{Montambaux2016}%
  \BibitemOpen
  \bibfield  {author} {\bibinfo {author} {\bibfnamefont {P.}~\bibnamefont
  {Adroguer}}, \bibinfo {author} {\bibfnamefont {D.}~\bibnamefont
  {Carpentier}}, \bibinfo {author} {\bibfnamefont {G.}~\bibnamefont
  {Montambaux}}, \ and\ \bibinfo {author} {\bibfnamefont {E.}~\bibnamefont
  {Orignac}},\ }\href {\doibase 10.1103/PhysRevB.93.125113} {\bibfield
  {journal} {\bibinfo  {journal} {Phys. Rev. B}\ }\textbf {\bibinfo {volume}
  {93}},\ \bibinfo {pages} {125113} (\bibinfo {year} {2016})}\BibitemShut
  {NoStop}%
\bibitem [{\citenamefont {Bellec}\ \emph {et~al.}(2013)\citenamefont {Bellec},
  \citenamefont {Kuhl}, \citenamefont {Montambaux},\ and\ \citenamefont
  {Mortessagne}}]{Bellec2013}%
  \BibitemOpen
  \bibfield  {author} {\bibinfo {author} {\bibfnamefont {M.}~\bibnamefont
  {Bellec}}, \bibinfo {author} {\bibfnamefont {U.}~\bibnamefont {Kuhl}},
  \bibinfo {author} {\bibfnamefont {G.}~\bibnamefont {Montambaux}}, \ and\
  \bibinfo {author} {\bibfnamefont {F.}~\bibnamefont {Mortessagne}},\ }\href
  {\doibase 10.1103/PhysRevLett.110.033902} {\bibfield  {journal} {\bibinfo
  {journal} {Phys. Rev. Lett.}\ }\textbf {\bibinfo {volume} {110}},\ \bibinfo
  {pages} {033902} (\bibinfo {year} {2013})}\BibitemShut {NoStop}%
\bibitem [{\citenamefont {Lim}\ \emph {et~al.}(2012)\citenamefont {Lim},
  \citenamefont {Fuchs},\ and\ \citenamefont {Montambaux}}]{Lim2012}%
  \BibitemOpen
  \bibfield  {author} {\bibinfo {author} {\bibfnamefont {L.-K.}\ \bibnamefont
  {Lim}}, \bibinfo {author} {\bibfnamefont {J.-N.}\ \bibnamefont {Fuchs}}, \
  and\ \bibinfo {author} {\bibfnamefont {G.}~\bibnamefont {Montambaux}},\
  }\href {\doibase 10.1103/PhysRevLett.108.175303} {\bibfield  {journal}
  {\bibinfo  {journal} {Phys. Rev. Lett.}\ }\textbf {\bibinfo {volume} {108}},\
  \bibinfo {pages} {175303} (\bibinfo {year} {2012})}\BibitemShut {NoStop}%
\bibitem [{\citenamefont {Banerjee}\ \emph {et~al.}(2009)\citenamefont
  {Banerjee}, \citenamefont {Singh}, \citenamefont {Pardo},\ and\ \citenamefont
  {Pickett}}]{Banarjee2009}%
  \BibitemOpen
  \bibfield  {author} {\bibinfo {author} {\bibfnamefont {S.}~\bibnamefont
  {Banerjee}}, \bibinfo {author} {\bibfnamefont {R.~R.~P.}\ \bibnamefont
  {Singh}}, \bibinfo {author} {\bibfnamefont {V.}~\bibnamefont {Pardo}}, \ and\
  \bibinfo {author} {\bibfnamefont {W.~E.}\ \bibnamefont {Pickett}},\ }\href
  {\doibase 10.1103/PhysRevLett.103.016402} {\bibfield  {journal} {\bibinfo
  {journal} {Phys. Rev. Lett.}\ }\textbf {\bibinfo {volume} {103}},\ \bibinfo
  {pages} {016402} (\bibinfo {year} {2009})}\BibitemShut {NoStop}%
\bibitem [{\citenamefont {Banerjee}\ and\ \citenamefont
  {Pickett}(2012)}]{Banarjee2012}%
  \BibitemOpen
  \bibfield  {author} {\bibinfo {author} {\bibfnamefont {S.}~\bibnamefont
  {Banerjee}}\ and\ \bibinfo {author} {\bibfnamefont {W.~E.}\ \bibnamefont
  {Pickett}},\ }\href {\doibase 10.1103/PhysRevB.86.075124} {\bibfield
  {journal} {\bibinfo  {journal} {Phys. Rev. B}\ }\textbf {\bibinfo {volume}
  {86}},\ \bibinfo {pages} {075124} (\bibinfo {year} {2012})}\BibitemShut
  {NoStop}%
\bibitem [{\citenamefont {Amorim}\ \emph {et~al.}(2016)\citenamefont {Amorim},
  \citenamefont {Cortijo}, \citenamefont {de~Juan}, \citenamefont {Grushin},
  \citenamefont {Guinea}, \citenamefont {{Guti{\'e}rrez-Rubio}}, \citenamefont
  {Ochoa}, \citenamefont {Parente}, \citenamefont {{Rold{\'a}n}}, \citenamefont
  {San-Jose}, \citenamefont {Schiefele}, \citenamefont {Sturla},\ and\
  \citenamefont {Vozmediano}}]{Maria2016}%
  \BibitemOpen
  \bibfield  {author} {\bibinfo {author} {\bibfnamefont {B.}~\bibnamefont
  {Amorim}}, \bibinfo {author} {\bibfnamefont {A.}~\bibnamefont {Cortijo}},
  \bibinfo {author} {\bibfnamefont {F.}~\bibnamefont {de~Juan}}, \bibinfo
  {author} {\bibfnamefont {A.}~\bibnamefont {Grushin}}, \bibinfo {author}
  {\bibfnamefont {F.}~\bibnamefont {Guinea}}, \bibinfo {author} {\bibfnamefont
  {A.}~\bibnamefont {{Guti{\'e}rrez-Rubio}}}, \bibinfo {author} {\bibfnamefont
  {H.}~\bibnamefont {Ochoa}}, \bibinfo {author} {\bibfnamefont
  {V.}~\bibnamefont {Parente}}, \bibinfo {author} {\bibfnamefont
  {R.}~\bibnamefont {{Rold{\'a}n}}}, \bibinfo {author} {\bibfnamefont
  {P.}~\bibnamefont {San-Jose}}, \bibinfo {author} {\bibfnamefont
  {J.}~\bibnamefont {Schiefele}}, \bibinfo {author} {\bibfnamefont
  {M.}~\bibnamefont {Sturla}}, \ and\ \bibinfo {author} {\bibfnamefont
  {M.}~\bibnamefont {Vozmediano}},\ }\href {\doibase
  http://dx.doi.org/10.1016/j.physrep.2015.12.006} {\bibfield  {journal}
  {\bibinfo  {journal} {Physics Reports}\ }\textbf {\bibinfo {volume} {617}},\
  \bibinfo {pages} {1 } (\bibinfo {year} {2016})}\BibitemShut {NoStop}%
\bibitem [{\citenamefont {Rodin}\ \emph {et~al.}(2014)\citenamefont {Rodin},
  \citenamefont {Carvalho},\ and\ \citenamefont
  {Castro~Neto}}]{PhysRevLett.112.176801}%
  \BibitemOpen
  \bibfield  {author} {\bibinfo {author} {\bibfnamefont {A.~S.}\ \bibnamefont
  {Rodin}}, \bibinfo {author} {\bibfnamefont {A.}~\bibnamefont {Carvalho}}, \
  and\ \bibinfo {author} {\bibfnamefont {A.~H.}\ \bibnamefont {Castro~Neto}},\
  }\href {\doibase 10.1103/PhysRevLett.112.176801} {\bibfield  {journal}
  {\bibinfo  {journal} {Phys. Rev. Lett.}\ }\textbf {\bibinfo {volume} {112}},\
  \bibinfo {pages} {176801} (\bibinfo {year} {2014})}\BibitemShut {NoStop}%
\bibitem [{\citenamefont {Kim}\ \emph {et~al.}(2015)\citenamefont {Kim},
  \citenamefont {Baik}, \citenamefont {Ryu}, \citenamefont {Sohn},
  \citenamefont {Park}, \citenamefont {Park}, \citenamefont {Denlinger},
  \citenamefont {Yi}, \citenamefont {Choi},\ and\ \citenamefont
  {Kim}}]{Kim723}%
  \BibitemOpen
  \bibfield  {author} {\bibinfo {author} {\bibfnamefont {J.}~\bibnamefont
  {Kim}}, \bibinfo {author} {\bibfnamefont {S.~S.}\ \bibnamefont {Baik}},
  \bibinfo {author} {\bibfnamefont {S.~H.}\ \bibnamefont {Ryu}}, \bibinfo
  {author} {\bibfnamefont {Y.}~\bibnamefont {Sohn}}, \bibinfo {author}
  {\bibfnamefont {S.}~\bibnamefont {Park}}, \bibinfo {author} {\bibfnamefont
  {B.-G.}\ \bibnamefont {Park}}, \bibinfo {author} {\bibfnamefont
  {J.}~\bibnamefont {Denlinger}}, \bibinfo {author} {\bibfnamefont
  {Y.}~\bibnamefont {Yi}}, \bibinfo {author} {\bibfnamefont {H.~J.}\
  \bibnamefont {Choi}}, \ and\ \bibinfo {author} {\bibfnamefont {K.~S.}\
  \bibnamefont {Kim}},\ }\href {\doibase 10.1126/science.aaa6486} {\bibfield
  {journal} {\bibinfo  {journal} {Science}\ }\textbf {\bibinfo {volume}
  {349}},\ \bibinfo {pages} {723} (\bibinfo {year} {2015})}\BibitemShut
  {NoStop}%
\bibitem [{\citenamefont {Katayama}\ \emph {et~al.}(2006)\citenamefont
  {Katayama}, \citenamefont {Kobayashi},\ and\ \citenamefont
  {Suzumura}}]{doi:10.1143/JPSJ.75.054705}%
  \BibitemOpen
  \bibfield  {author} {\bibinfo {author} {\bibfnamefont {S.}~\bibnamefont
  {Katayama}}, \bibinfo {author} {\bibfnamefont {A.}~\bibnamefont {Kobayashi}},
  \ and\ \bibinfo {author} {\bibfnamefont {Y.}~\bibnamefont {Suzumura}},\
  }\href {\doibase 10.1143/JPSJ.75.054705} {\bibfield  {journal} {\bibinfo
  {journal} {Journal of the Physical Society of Japan}\ }\textbf {\bibinfo
  {volume} {75}},\ \bibinfo {pages} {054705} (\bibinfo {year}
  {2006})}\BibitemShut {NoStop}%
\bibitem [{\citenamefont {Pardo}\ and\ \citenamefont
  {Pickett}(2009)}]{PhysRevLett.102.166803}%
  \BibitemOpen
  \bibfield  {author} {\bibinfo {author} {\bibfnamefont {V.}~\bibnamefont
  {Pardo}}\ and\ \bibinfo {author} {\bibfnamefont {W.~E.}\ \bibnamefont
  {Pickett}},\ }\href {\doibase 10.1103/PhysRevLett.102.166803} {\bibfield
  {journal} {\bibinfo  {journal} {Phys. Rev. Lett.}\ }\textbf {\bibinfo
  {volume} {102}},\ \bibinfo {pages} {166803} (\bibinfo {year}
  {2009})}\BibitemShut {NoStop}%
\bibitem [{\citenamefont {Huang}\ \emph {et~al.}(2015)\citenamefont {Huang},
  \citenamefont {Liu}, \citenamefont {Zhang}, \citenamefont {Duan},\ and\
  \citenamefont {Vanderbilt}}]{PhysRevB.92.161115}%
  \BibitemOpen
  \bibfield  {author} {\bibinfo {author} {\bibfnamefont {H.}~\bibnamefont
  {Huang}}, \bibinfo {author} {\bibfnamefont {Z.}~\bibnamefont {Liu}}, \bibinfo
  {author} {\bibfnamefont {H.}~\bibnamefont {Zhang}}, \bibinfo {author}
  {\bibfnamefont {W.}~\bibnamefont {Duan}}, \ and\ \bibinfo {author}
  {\bibfnamefont {D.}~\bibnamefont {Vanderbilt}},\ }\href {\doibase
  10.1103/PhysRevB.92.161115} {\bibfield  {journal} {\bibinfo  {journal} {Phys.
  Rev. B}\ }\textbf {\bibinfo {volume} {92}},\ \bibinfo {pages} {161115}
  (\bibinfo {year} {2015})}\BibitemShut {NoStop}%
\bibitem [{\citenamefont {Rechtsman}\ \emph {et~al.}(2013)\citenamefont
  {Rechtsman}, \citenamefont {Plotnik}, \citenamefont {Zeuner}, \citenamefont
  {Song}, \citenamefont {Chen}, \citenamefont {Szameit},\ and\ \citenamefont
  {Segev}}]{Photonic2013}%
  \BibitemOpen
  \bibfield  {author} {\bibinfo {author} {\bibfnamefont {M.~C.}\ \bibnamefont
  {Rechtsman}}, \bibinfo {author} {\bibfnamefont {Y.}~\bibnamefont {Plotnik}},
  \bibinfo {author} {\bibfnamefont {J.~M.}\ \bibnamefont {Zeuner}}, \bibinfo
  {author} {\bibfnamefont {D.}~\bibnamefont {Song}}, \bibinfo {author}
  {\bibfnamefont {Z.}~\bibnamefont {Chen}}, \bibinfo {author} {\bibfnamefont
  {A.}~\bibnamefont {Szameit}}, \ and\ \bibinfo {author} {\bibfnamefont
  {M.}~\bibnamefont {Segev}},\ }\href {\doibase 10.1103/PhysRevLett.111.103901}
  {\bibfield  {journal} {\bibinfo  {journal} {Phys. Rev. Lett.}\ }\textbf
  {\bibinfo {volume} {111}},\ \bibinfo {pages} {103901} (\bibinfo {year}
  {2013})}\BibitemShut {NoStop}%
\bibitem [{\citenamefont {Polini}\ \emph {et~al.}(2013)\citenamefont {Polini},
  \citenamefont {Guinea}, \citenamefont {Lewenstein}, \citenamefont
  {Manoharan},\ and\ \citenamefont {Pellegrini}}]{Polini2013}%
  \BibitemOpen
  \bibfield  {author} {\bibinfo {author} {\bibfnamefont {M.}~\bibnamefont
  {Polini}}, \bibinfo {author} {\bibfnamefont {F.}~\bibnamefont {Guinea}},
  \bibinfo {author} {\bibfnamefont {M.}~\bibnamefont {Lewenstein}}, \bibinfo
  {author} {\bibfnamefont {H.~C.}\ \bibnamefont {Manoharan}}, \ and\ \bibinfo
  {author} {\bibfnamefont {V.}~\bibnamefont {Pellegrini}},\ }\href
  {http://dx.doi.org/10.1038/nnano.2013.161} {\bibfield  {journal} {\bibinfo
  {journal} {Nature Nanotechnology}\ }\textbf {\bibinfo {volume} {8}},\
  \bibinfo {pages} {625} (\bibinfo {year} {2013})}\BibitemShut {NoStop}%
\bibitem [{\citenamefont {Pereira}\ \emph {et~al.}(2009)\citenamefont
  {Pereira}, \citenamefont {Castro~Neto},\ and\ \citenamefont
  {Peres}}]{Pereira2009}%
  \BibitemOpen
  \bibfield  {author} {\bibinfo {author} {\bibfnamefont {V.~M.}\ \bibnamefont
  {Pereira}}, \bibinfo {author} {\bibfnamefont {A.~H.}\ \bibnamefont
  {Castro~Neto}}, \ and\ \bibinfo {author} {\bibfnamefont {N.~M.~R.}\
  \bibnamefont {Peres}},\ }\href {\doibase 10.1103/PhysRevB.80.045401}
  {\bibfield  {journal} {\bibinfo  {journal} {Phys. Rev. B}\ }\textbf {\bibinfo
  {volume} {80}},\ \bibinfo {pages} {045401} (\bibinfo {year}
  {2009})}\BibitemShut {NoStop}%
\bibitem [{\citenamefont {Choi}\ \emph {et~al.}(2010)\citenamefont {Choi},
  \citenamefont {Jhi},\ and\ \citenamefont {Son}}]{Choi2010}%
  \BibitemOpen
  \bibfield  {author} {\bibinfo {author} {\bibfnamefont {S.-M.}\ \bibnamefont
  {Choi}}, \bibinfo {author} {\bibfnamefont {S.-H.}\ \bibnamefont {Jhi}}, \
  and\ \bibinfo {author} {\bibfnamefont {Y.-W.}\ \bibnamefont {Son}},\ }\href
  {\doibase 10.1103/PhysRevB.81.081407} {\bibfield  {journal} {\bibinfo
  {journal} {Phys. Rev. B}\ }\textbf {\bibinfo {volume} {81}},\ \bibinfo
  {pages} {081407} (\bibinfo {year} {2010})}\BibitemShut {NoStop}%
\bibitem [{\citenamefont {D\'ora}\ \emph {et~al.}(2013)\citenamefont {D\'ora},
  \citenamefont {Herbut},\ and\ \citenamefont {Moessner}}]{Dora2013}%
  \BibitemOpen
  \bibfield  {author} {\bibinfo {author} {\bibfnamefont {B.}~\bibnamefont
  {D\'ora}}, \bibinfo {author} {\bibfnamefont {I.~F.}\ \bibnamefont {Herbut}},
  \ and\ \bibinfo {author} {\bibfnamefont {R.}~\bibnamefont {Moessner}},\
  }\href {\doibase 10.1103/PhysRevB.88.075126} {\bibfield  {journal} {\bibinfo
  {journal} {Phys. Rev. B}\ }\textbf {\bibinfo {volume} {88}},\ \bibinfo
  {pages} {075126} (\bibinfo {year} {2013})}\BibitemShut {NoStop}%
\bibitem [{\citenamefont {Uryszek}\ \emph {et~al.}(2019)\citenamefont
  {Uryszek}, \citenamefont {Christou}, \citenamefont {Jaefari}, \citenamefont
  {Kr\"uger},\ and\ \citenamefont {Uchoa}}]{Uryszek2019}%
  \BibitemOpen
  \bibfield  {author} {\bibinfo {author} {\bibfnamefont {M.~D.}\ \bibnamefont
  {Uryszek}}, \bibinfo {author} {\bibfnamefont {E.}~\bibnamefont {Christou}},
  \bibinfo {author} {\bibfnamefont {A.}~\bibnamefont {Jaefari}}, \bibinfo
  {author} {\bibfnamefont {F.}~\bibnamefont {Kr\"uger}}, \ and\ \bibinfo
  {author} {\bibfnamefont {B.}~\bibnamefont {Uchoa}},\ }\href {\doibase
  10.1103/PhysRevB.100.155101} {\bibfield  {journal} {\bibinfo  {journal}
  {Phys. Rev. B}\ }\textbf {\bibinfo {volume} {100}},\ \bibinfo {pages}
  {155101} (\bibinfo {year} {2019})}\BibitemShut {NoStop}%
\bibitem [{\citenamefont {Uchoa}\ and\ \citenamefont
  {Seo}(2017)}]{PhysRevB.96.220503}%
  \BibitemOpen
  \bibfield  {author} {\bibinfo {author} {\bibfnamefont {B.}~\bibnamefont
  {Uchoa}}\ and\ \bibinfo {author} {\bibfnamefont {K.}~\bibnamefont {Seo}},\
  }\href {\doibase 10.1103/PhysRevB.96.220503} {\bibfield  {journal} {\bibinfo
  {journal} {Phys. Rev. B}\ }\textbf {\bibinfo {volume} {96}},\ \bibinfo
  {pages} {220503} (\bibinfo {year} {2017})}\BibitemShut {NoStop}%
\bibitem [{\citenamefont {Roy}\ and\ \citenamefont
  {Foster}(2018)}]{Foster2018}%
  \BibitemOpen
  \bibfield  {author} {\bibinfo {author} {\bibfnamefont {B.}~\bibnamefont
  {Roy}}\ and\ \bibinfo {author} {\bibfnamefont {M.~S.}\ \bibnamefont
  {Foster}},\ }\href {\doibase 10.1103/PhysRevX.8.011049} {\bibfield  {journal}
  {\bibinfo  {journal} {Phys. Rev. X}\ }\textbf {\bibinfo {volume} {8}},\
  \bibinfo {pages} {011049} (\bibinfo {year} {2018})}\BibitemShut {NoStop}%
\bibitem [{\citenamefont {Isobe}\ \emph {et~al.}(2016)\citenamefont {Isobe},
  \citenamefont {Yang}, \citenamefont {Chubukov}, \citenamefont {Schmalian},\
  and\ \citenamefont {Nagaosa}}]{Isobe2016}%
  \BibitemOpen
  \bibfield  {author} {\bibinfo {author} {\bibfnamefont {H.}~\bibnamefont
  {Isobe}}, \bibinfo {author} {\bibfnamefont {B.-J.}\ \bibnamefont {Yang}},
  \bibinfo {author} {\bibfnamefont {A.}~\bibnamefont {Chubukov}}, \bibinfo
  {author} {\bibfnamefont {J.}~\bibnamefont {Schmalian}}, \ and\ \bibinfo
  {author} {\bibfnamefont {N.}~\bibnamefont {Nagaosa}},\ }\href {\doibase
  10.1103/PhysRevLett.116.076803} {\bibfield  {journal} {\bibinfo  {journal}
  {Phys. Rev. Lett.}\ }\textbf {\bibinfo {volume} {116}},\ \bibinfo {pages}
  {076803} (\bibinfo {year} {2016})}\BibitemShut {NoStop}%
\bibitem [{\citenamefont {Cho}\ and\ \citenamefont {Moon}(2016)}]{Cho2016}%
  \BibitemOpen
  \bibfield  {author} {\bibinfo {author} {\bibfnamefont {G.~Y.}\ \bibnamefont
  {Cho}}\ and\ \bibinfo {author} {\bibfnamefont {E.-G.}\ \bibnamefont {Moon}},\
  }\href@noop {} {\bibfield  {journal} {\bibinfo  {journal} {Sci. Rep.}\
  }\textbf {\bibinfo {volume} {6}},\ \bibinfo {pages} {19198} (\bibinfo {year}
  {2016})}\BibitemShut {NoStop}%
\bibitem [{\citenamefont {Link}\ \emph {et~al.}(2018)\citenamefont {Link},
  \citenamefont {Narozhny}, \citenamefont {Kiselev},\ and\ \citenamefont
  {Schmalian}}]{Link2018}%
  \BibitemOpen
  \bibfield  {author} {\bibinfo {author} {\bibfnamefont {J.~M.}\ \bibnamefont
  {Link}}, \bibinfo {author} {\bibfnamefont {B.~N.}\ \bibnamefont {Narozhny}},
  \bibinfo {author} {\bibfnamefont {E.~I.}\ \bibnamefont {Kiselev}}, \ and\
  \bibinfo {author} {\bibfnamefont {J.}~\bibnamefont {Schmalian}},\ }\href
  {\doibase 10.1103/PhysRevLett.120.196801} {\bibfield  {journal} {\bibinfo
  {journal} {Phys. Rev. Lett.}\ }\textbf {\bibinfo {volume} {120}},\ \bibinfo
  {pages} {196801} (\bibinfo {year} {2018})}\BibitemShut {NoStop}%
\bibitem [{\citenamefont {Son}(2007)}]{Son2007}%
  \BibitemOpen
  \bibfield  {author} {\bibinfo {author} {\bibfnamefont {D.~T.}\ \bibnamefont
  {Son}},\ }\href {\doibase 10.1103/PhysRevB.75.235423} {\bibfield  {journal}
  {\bibinfo  {journal} {Phys. Rev. B}\ }\textbf {\bibinfo {volume} {75}},\
  \bibinfo {pages} {235423} (\bibinfo {year} {2007})}\BibitemShut {NoStop}%
\bibitem [{\citenamefont {Kotov}\ \emph {et~al.}(2009)\citenamefont {Kotov},
  \citenamefont {Uchoa},\ and\ \citenamefont {Castro~Neto}}]{Kotov2009}%
  \BibitemOpen
  \bibfield  {author} {\bibinfo {author} {\bibfnamefont {V.~N.}\ \bibnamefont
  {Kotov}}, \bibinfo {author} {\bibfnamefont {B.}~\bibnamefont {Uchoa}}, \ and\
  \bibinfo {author} {\bibfnamefont {A.~H.}\ \bibnamefont {Castro~Neto}},\
  }\href {\doibase 10.1103/PhysRevB.80.165424} {\bibfield  {journal} {\bibinfo
  {journal} {Phys. Rev. B}\ }\textbf {\bibinfo {volume} {80}},\ \bibinfo
  {pages} {165424} (\bibinfo {year} {2009})}\BibitemShut {NoStop}%
\bibitem [{\citenamefont {Kotov}\ \emph {et~al.}(2012)\citenamefont {Kotov},
  \citenamefont {Uchoa}, \citenamefont {Pereira}, \citenamefont {Guinea},\ and\
  \citenamefont {Castro~Neto}}]{Kotov2012}%
  \BibitemOpen
  \bibfield  {author} {\bibinfo {author} {\bibfnamefont {V.~N.}\ \bibnamefont
  {Kotov}}, \bibinfo {author} {\bibfnamefont {B.}~\bibnamefont {Uchoa}},
  \bibinfo {author} {\bibfnamefont {V.~M.}\ \bibnamefont {Pereira}}, \bibinfo
  {author} {\bibfnamefont {F.}~\bibnamefont {Guinea}}, \ and\ \bibinfo {author}
  {\bibfnamefont {A.~H.}\ \bibnamefont {Castro~Neto}},\ }\href {\doibase
  10.1103/RevModPhys.84.1067} {\bibfield  {journal} {\bibinfo  {journal} {Rev.
  Mod. Phys.}\ }\textbf {\bibinfo {volume} {84}},\ \bibinfo {pages} {1067}
  (\bibinfo {year} {2012})}\BibitemShut {NoStop}%
\bibitem [{\citenamefont {Kovtun}\ \emph {et~al.}(2005)\citenamefont {Kovtun},
  \citenamefont {Son},\ and\ \citenamefont
  {Starinets}}]{PhysRevLett.94.111601}%
  \BibitemOpen
  \bibfield  {author} {\bibinfo {author} {\bibfnamefont {P.~K.}\ \bibnamefont
  {Kovtun}}, \bibinfo {author} {\bibfnamefont {D.~T.}\ \bibnamefont {Son}}, \
  and\ \bibinfo {author} {\bibfnamefont {A.~O.}\ \bibnamefont {Starinets}},\
  }\href {\doibase 10.1103/PhysRevLett.94.111601} {\bibfield  {journal}
  {\bibinfo  {journal} {Phys. Rev. Lett.}\ }\textbf {\bibinfo {volume} {94}},\
  \bibinfo {pages} {111601} (\bibinfo {year} {2005})}\BibitemShut {NoStop}%
\bibitem [{\citenamefont {M\"uller}\ \emph {et~al.}(2009)\citenamefont
  {M\"uller}, \citenamefont {Schmalian},\ and\ \citenamefont
  {Fritz}}]{PhysRevLett.103.025301}%
  \BibitemOpen
  \bibfield  {author} {\bibinfo {author} {\bibfnamefont {M.}~\bibnamefont
  {M\"uller}}, \bibinfo {author} {\bibfnamefont {J.}~\bibnamefont {Schmalian}},
  \ and\ \bibinfo {author} {\bibfnamefont {L.}~\bibnamefont {Fritz}},\ }\href
  {\doibase 10.1103/PhysRevLett.103.025301} {\bibfield  {journal} {\bibinfo
  {journal} {Phys. Rev. Lett.}\ }\textbf {\bibinfo {volume} {103}},\ \bibinfo
  {pages} {025301} (\bibinfo {year} {2009})}\BibitemShut {NoStop}%
\bibitem [{\citenamefont {Herbut}(2006)}]{Herbut2006}%
  \BibitemOpen
  \bibfield  {author} {\bibinfo {author} {\bibfnamefont {I.~F.}\ \bibnamefont
  {Herbut}},\ }\href {\doibase 10.1103/PhysRevLett.97.146401} {\bibfield
  {journal} {\bibinfo  {journal} {Phys. Rev. Lett.}\ }\textbf {\bibinfo
  {volume} {97}},\ \bibinfo {pages} {146401} (\bibinfo {year}
  {2006})}\BibitemShut {NoStop}%
\bibitem [{\citenamefont {Ye}\ and\ \citenamefont {Sachdev}(1998)}]{Ye1998}%
  \BibitemOpen
  \bibfield  {author} {\bibinfo {author} {\bibfnamefont {J.}~\bibnamefont
  {Ye}}\ and\ \bibinfo {author} {\bibfnamefont {S.}~\bibnamefont {Sachdev}},\
  }\href {\doibase 10.1103/PhysRevLett.80.5409} {\bibfield  {journal} {\bibinfo
   {journal} {Phys. Rev. Lett.}\ }\textbf {\bibinfo {volume} {80}},\ \bibinfo
  {pages} {5409} (\bibinfo {year} {1998})}\BibitemShut {NoStop}%
\bibitem [{\citenamefont {Wang}\ \emph {et~al.}(2017)\citenamefont {Wang},
  \citenamefont {Liu},\ and\ \citenamefont {Zhang}}]{Wang17}%
  \BibitemOpen
  \bibfield  {author} {\bibinfo {author} {\bibfnamefont {J.-R.}\ \bibnamefont
  {Wang}}, \bibinfo {author} {\bibfnamefont {G.-Z.}\ \bibnamefont {Liu}}, \
  and\ \bibinfo {author} {\bibfnamefont {C.-J.}\ \bibnamefont {Zhang}},\ }\href
  {\doibase 10.1103/PhysRevB.95.075129} {\bibfield  {journal} {\bibinfo
  {journal} {Phys. Rev. B}\ }\textbf {\bibinfo {volume} {95}},\ \bibinfo
  {pages} {075129} (\bibinfo {year} {2017})}\BibitemShut {NoStop}%
\bibitem [{\citenamefont {Vafek}\ \emph {et~al.}(2002)\citenamefont {Vafek},
  \citenamefont {Te\ifmmode \check{s}\else
  \v{s}\fi{}anovi\ifmmode~\acute{c}\else \'{c}\fi{}},\ and\ \citenamefont
  {Franz}}]{Vafek2002}%
  \BibitemOpen
  \bibfield  {author} {\bibinfo {author} {\bibfnamefont {O.}~\bibnamefont
  {Vafek}}, \bibinfo {author} {\bibfnamefont {Z.}~\bibnamefont {Te\ifmmode
  \check{s}\else \v{s}\fi{}anovi\ifmmode~\acute{c}\else \'{c}\fi{}}}, \ and\
  \bibinfo {author} {\bibfnamefont {M.}~\bibnamefont {Franz}},\ }\href
  {\doibase 10.1103/PhysRevLett.89.157003} {\bibfield  {journal} {\bibinfo
  {journal} {Phys. Rev. Lett.}\ }\textbf {\bibinfo {volume} {89}},\ \bibinfo
  {pages} {157003} (\bibinfo {year} {2002})}\BibitemShut {NoStop}%
\bibitem [{\citenamefont {Martin}\ \emph {et~al.}(2008)\citenamefont {Martin},
  \citenamefont {Akerman}, \citenamefont {Ulbricht}, \citenamefont {Lohmann},
  \citenamefont {Smet}, \citenamefont {{von Klitzing}},\ and\ \citenamefont
  {Yacoby}}]{Yacoby08}%
  \BibitemOpen
  \bibfield  {author} {\bibinfo {author} {\bibfnamefont {J.}~\bibnamefont
  {Martin}}, \bibinfo {author} {\bibfnamefont {N.}~\bibnamefont {Akerman}},
  \bibinfo {author} {\bibfnamefont {G.}~\bibnamefont {Ulbricht}}, \bibinfo
  {author} {\bibfnamefont {T.}~\bibnamefont {Lohmann}}, \bibinfo {author}
  {\bibfnamefont {J.~H.}\ \bibnamefont {Smet}}, \bibinfo {author}
  {\bibfnamefont {K.}~\bibnamefont {{von Klitzing}}}, \ and\ \bibinfo {author}
  {\bibfnamefont {A.}~\bibnamefont {Yacoby}},\ }\href@noop {} {\bibfield
  {journal} {\bibinfo  {journal} {Nature Physics}\ }\textbf {\bibinfo {volume}
  {4}},\ \bibinfo {pages} {144} (\bibinfo {year} {2008})}\BibitemShut {NoStop}%
\bibitem [{\citenamefont {Yu}\ \emph {et~al.}(2013)\citenamefont {Yu},
  \citenamefont {Jalil}, \citenamefont {Belle}, \citenamefont {Mayorov},
  \citenamefont {Blake}, \citenamefont {Schedin}, \citenamefont {Morozov},
  \citenamefont {Ponomarenko}, \citenamefont {Chiappini}, \citenamefont
  {Wiedmann}, \citenamefont {Zeitler}, \citenamefont {Katsnelson},
  \citenamefont {Geim}, \citenamefont {Novoselov},\ and\ \citenamefont
  {Elias}}]{Geim13}%
  \BibitemOpen
  \bibfield  {author} {\bibinfo {author} {\bibfnamefont {G.~L.}\ \bibnamefont
  {Yu}}, \bibinfo {author} {\bibfnamefont {R.}~\bibnamefont {Jalil}}, \bibinfo
  {author} {\bibfnamefont {B.}~\bibnamefont {Belle}}, \bibinfo {author}
  {\bibfnamefont {A.~S.}\ \bibnamefont {Mayorov}}, \bibinfo {author}
  {\bibfnamefont {P.}~\bibnamefont {Blake}}, \bibinfo {author} {\bibfnamefont
  {F.}~\bibnamefont {Schedin}}, \bibinfo {author} {\bibfnamefont {S.~V.}\
  \bibnamefont {Morozov}}, \bibinfo {author} {\bibfnamefont {L.~A.}\
  \bibnamefont {Ponomarenko}}, \bibinfo {author} {\bibfnamefont
  {F.}~\bibnamefont {Chiappini}}, \bibinfo {author} {\bibfnamefont
  {S.}~\bibnamefont {Wiedmann}}, \bibinfo {author} {\bibfnamefont
  {U.}~\bibnamefont {Zeitler}}, \bibinfo {author} {\bibfnamefont {M.~I.}\
  \bibnamefont {Katsnelson}}, \bibinfo {author} {\bibfnamefont {A.~K.}\
  \bibnamefont {Geim}}, \bibinfo {author} {\bibfnamefont {K.~S.}\ \bibnamefont
  {Novoselov}}, \ and\ \bibinfo {author} {\bibfnamefont {D.~C.}\ \bibnamefont
  {Elias}},\ }\href {\doibase 10.1073/pnas.1300599110} {\bibfield  {journal}
  {\bibinfo  {journal} {Proceedings of the National Academy of Sciences}\
  }\textbf {\bibinfo {volume} {110}},\ \bibinfo {pages} {3282} (\bibinfo {year}
  {2013})}\BibitemShut {NoStop}%
\bibitem [{\citenamefont {Sheehy}\ and\ \citenamefont
  {Schmalian}(2007)}]{Sheehy07}%
  \BibitemOpen
  \bibfield  {author} {\bibinfo {author} {\bibfnamefont {D.~E.}\ \bibnamefont
  {Sheehy}}\ and\ \bibinfo {author} {\bibfnamefont {J.}~\bibnamefont
  {Schmalian}},\ }\href@noop {} {\bibfield  {journal} {\bibinfo  {journal}
  {\prl}\ }\textbf {\bibinfo {volume} {99}},\ \bibinfo {pages} {226803}
  (\bibinfo {year} {2007})}\BibitemShut {NoStop}%
\end{thebibliography}%

\end{document}